\documentclass[final,1p,times]{elsarticle}

\usepackage{graphics}

\usepackage{amssymb}
\usepackage{subeqn}
\usepackage{color}
\usepackage{tabularx}
\usepackage{multirow}
\UseRawInputEncoding
\biboptions{sort&compress}

\journal{}

\begin{document}

\begin{frontmatter}

\title{Multiple-distribution-function lattice Boltzmann method for convection-diffusion-system based incompressible Navier-Stokes equations}

\author{Zhenhua Chai\fnref{a,b}}\ead{hustczh@hust.edu.cn}
\author[a,b]{Baochang Shi\corref{cor1}}\cortext[cor1]{Corresponding author at: School of Mathematics and Statistics, Huazhong University of Science and Technology, Wuhan, 430074, China. Tel./fax: +86 27 8754 3231.
\hspace*{20pt}}\ead{shibc@hust.edu.cn}
\author{Chengjie Zhan\fnref{a}}
\address[a]{School of Mathematics and Statistics, Huazhong University of Science and Technology, Wuhan, 430074, China}
\address[b]{Hubei Key Laboratory of Engineering Modeling and Scientific Computing, Huazhong University of Science and Technology, Wuhan 430074, China}

\begin{abstract}
In this paper, a multiple-distribution-function lattice Boltzmann method (MDF-LBM) with multiple-relaxation-time model is proposed for incompressible Navier-Stokes equations (NSEs) which are considered as the coupled convection-diffusion equations (CDEs). Through direct Taylor expansion analysis, we show that the Navier-Stokes equations can be recovered correctly from the present MDF-LBM, and additionally, it is also found that the velocity and pressure can be directly computed through the zero and first-order moments of distribution function. Then in the framework of present MDF-LBM, we develop a locally computational scheme for the velocity gradient where the first-order moment of the non-equilibrium distribution is used, this scheme is also extended to calculate the velocity divergence, strain rate tensor, shear stress and vorticity. Finally, we also conduct some simulations to test the MDF-LBM, and find that the numerical results not only agree with some available analytical and numerical solutions, but also have a second-order convergence rate in space.
\end{abstract}


\end{frontmatter}

%
%

\section{Introduction}
In the past three decades, the lattice Boltzmann method (LBM), as a discrete numerical approach to the Boltzmann equation, has attained a great success in the study of the complex fluid flows (see some review articles \cite{Benzi1992, Chen1998, Aidun2010, Xu2017, Wang2019} and monographs \cite{Wolf-Gladrow2000, Succi2001, Guo2013, Huang2015, Kruger2017}), including the multicomponent and multiphase flows \cite{Chen1998, Wang2019, Guo2013, Huang2015, Kruger2017}, thermal flows \cite{Succi2001, Guo2013}, turbulent flows \cite{Benzi1992, Chen1998, Wolf-Gladrow2000, Succi2001}, particle suspensions \cite{Aidun2010}, microfluidics \cite{Aidun2010, Guo2013}, flows in porous media \cite{Chen1998, Xu2017, Succi2001}, to name but a few.

Recently, the vectorial lattice Boltzmann method (VLBM) based on vectorial kinetic model (e.g., the general BGK model \cite{Bouchut1999,Carfora2008,Bianchini2019}) has also been developed for the shallow water equations \cite{Dubois2014} and incompressible Navier-Stokes equations (NSEs) \cite{Zhao2021,Zhao2020a,Zhao2020b}. Actually, for the isothermal and incompressible fluid flows where the density is assumed to be a constant, the NSEs can be viewed as a convection-diffusion system consisted of ($d+1$) equations in $d$-dimensional space [named convection-diffusion-system based NSEs, see the following Eq. (\ref{eq:2-2})]. The main idea of the VLBM is to construct a single evolution equation of distribution function for each convection-diffusion equation (CDE). Like the double-distribution-function LBM for thermal fluid flows \cite{Shan1997,He1998,Guo2002}, the VLBM can also be considered as a special multiple-distribution-function lattice Boltzmann method (MDF-LBM), which would be used in the present work.

Compared to the popular scalar or single LBM for the NSEs, the MDF-LBM (or VLBM) for convection-diffusion-system based NSEs has some distinct features. The first is that one can use fewer discrete velocities to construct the MDF-LBM. For instance, if we consider the two-dimensional problems, the LBM with D2Q4 or D2Q5 (four or five discrete velocities in two-dimensional space) lattice model is enough for the CDEs \cite{Cui2016}, while in the single LBM for the NSEs, the D2Q9 lattice model is usually adopted since the high-order isotropy of the discrete velocities is needed \cite{Qian1992}. The second is that in the MDF-LBM for convection-diffusion-system based NSEs, it is more flexible and much easier to construct the equilibrium distribution function such that the CDEs can be recovered correctly. And the third is that in the MDF-LBM for convection-diffusion-system based NSEs, some physical variables, for example, the velocity gradient, velocity divergence, the strain rate tensor, the shear stress and the vorticity, can be calculated locally through the first-order moments of the non-equilibrium distribution function (see Section III for details), while in the commonly used single LBM for NSEs, the second-order moments of the non-equilibrium distribution function are needed to compute strain rate tensor and shear stress \cite{Artoli2004, Kruger2009, Chai2012, Yong2012}, and what is more, it seems more difficult to directly develop the local scheme for velocity gradient or vorticity \cite{Peng2017}.

Inspired by the VLBM for the NSEs \cite{Zhao2021,Zhao2020a,Zhao2020b}, in this work we developed a MDF-LBM for the convection-diffusion-system based incompressible NSEs. Compared to the previous work \cite{Zhao2021,Zhao2020a,Zhao2020b}, however, there are four main differences: (1) we focus on the MDF-LBM for incompressible NSEs, and the compressible effect in Refs. \cite{Zhao2021,Zhao2020a,Zhao2020b} is neglected; (2) we propose a special formula with the first-order moment of the distribution function for the pressure, which is not only different from those considered in the previous work \cite{Zhao2021,Zhao2020a,Zhao2020b}, but also more strict theoretically; (3) the formula for the pressure is consistent with the continuity equation, and thus the lattice Boltzmann model for the continuity equation can be omitted; (4) we develop some local schemes for velocity gradient, velocity divergence, stain rate tensor, shear stress and vorticity, which have not been presented and discussed in the previous works \cite{Zhao2021,Zhao2020a,Zhao2020b}.

The rest of the paper is organized as follows. In section 2, we develop a MDF-LBM for convection-diffusion-system based incompressible NSEs, and then the direct Taylor expansion of the present MDF-LBM is conducted in section 3. In section 4, we present some numerical results and discussion, and finally, some conclusions are given in section 5.

\section{Multiple-distribution-function lattice Boltzmann method for incompressible Navier-Stokes equations}

In this section, we would first write the incompressible NSEs as a coupled convection-diffusion system, and then present a MDF-LBM for the convection-diffusion-system based incompressible NSEs.

\subsection{The convection-diffusion-system based incompressible Navier-Stokes equations}
For incompressible fluid flows where the density $\rho$ is assumed to be a positive constant $\rho_0$, the NSEs can be written as \cite{Landau,Kundu2016}
\begin{subequations}\label{eq:2-1}
\begin{equation}
 \nabla\cdot\mathbf{u}=R,
\end{equation}
\begin{equation}
 \frac{\partial\mathbf{u}}{\partial t}+\nabla\cdot(\mathbf{u}\mathbf{u})=-\nabla P+\nabla\cdot(\nu\nabla\mathbf{u})+\mathbf{F},
\end{equation}
\end{subequations}
where $\mathbf{u}=(u_\alpha)_{\alpha=1-d}$ is the velocity in $d$-dimensional space, $R$ is a source term, $P$ is the pressure, $\nu$ is the kinematic viscosity and $\mathbf{F}=(F_\alpha$) is the external force. We would like to point out that above NSEs can also be reformulated as a coupled convection-diffusion system,
\begin{equation}\label{eq:2-2}
 \frac{\partial\bar{u}_\alpha}{\partial t}+\nabla\cdot(\bar{u}_\alpha\mathbf{u}+P\mathbf{E}_\alpha)=\nabla\cdot(\nu\nabla\bar{u}_\alpha)+\bar{F}_\alpha,\ \alpha=0, 1, \cdots, d
\end{equation}
where $\bar{\mathbf{u}}=(\bar{u}_{\alpha})$ with $\bar{u}_{0}=\rho_0$ and $\bar{u}_{\alpha}=u_{\alpha}\ (\alpha=1-d)$, $\bar{\mathbf{F}}=(\bar{F}_{\alpha})$ with $\bar{F}_{0}=R$ and $\bar{F}_{\alpha}=F_{\alpha} (\alpha=1-d)$, $\mathbf{E} _{0}=\mathbf{0}$ and $\mathbf{E}_{\alpha} (\alpha=1-d)$ is the unit vector in $d$-dimensional space.

It is clear that Eq. (\ref{eq:2-2}) is composed of ($d+1$) convection-diffusion equations (CDEs), and in the following, it is considered as the convection-diffusion-system based NSEs. We note that although the incompressible NSEs (\ref{eq:2-1}) are equivalent to Eq. (\ref{eq:2-2}) mathematically, while the latter is just a convection-diffusion system, and can be solved more efficiently in the LBM. Actually, in the LBM for the CDE, only the second-order moment isotropy is needed, which brings more flexibility in the development of the lattice Boltzmann (LB) models; while in the single LBM for NSEs, we need the fourth-order moment isotropy, which would give rise to more limitations in the design of the LB models.

\subsection{The multiple-distribution-function lattice Boltzmann method for the convection-diffusion-system based Navier-Stokes-equations}

In the LBM, the LB models can be classified into three basic kinds, i.e., the single-relaxation-time LB (SRT-LB) model or lattice BGK model \cite{Qian1992, Chen1992}, the two-relaxation-time LB (TRT-LB) model \cite{Ginzburg2005a, Ginzburg2008} and the multiple-relaxation-time LB (MRT-LB) model \cite{dHumieres1992, Lallemand2000}. In this work, we would consider the MRT-LB model for its generality and accuracy \cite{Pan2006, Luo2011, Chai2016a, Chai2016b}. In the past years, some different MRT-LB models were developed for the isotropic and anisotropic CDEs \cite{Chai2016a,Rasin2005,Yoshida2010,Chai2014,Zhang2019}. Recently, Chai and Shi proposed a unified framework for the modeling of the MRT-LB models for the NSEs and nonlinear CDEs \cite{Chai2020}. Following this work and inspired by the VLBM \cite{Dubois2014,Zhao2021,Zhao2020a,Zhao2020b}, The evolution equation of the MDF-LBM for the convection-diffusion-system based NSEs (\ref{eq:2-2}) can be written as
\begin{equation}\label{eq:2-3}
 f_{i, \alpha}(\mathbf{x}+\mathbf{c}_i \delta t,t+\delta t)=f_{i, \alpha}(\mathbf{x}, t)-\mathbf{\Lambda}_{ik} \big[f_{k, \alpha}(\mathbf{x}, t)-f_{k, \alpha}^{eq}(\mathbf{x}, t)\big]+\delta t \big[G_{i, \alpha}(\mathbf{x},t)+F_{i, \alpha}(\mathbf{x}, t)+\frac{\delta t}{2}\bar{D}_i F_{i, \alpha}(\mathbf{x}, t)\big],
\end{equation}
where $f_{i, \alpha}(\mathbf{x}, t)$ is the distribution function corresponding to the variable $\bar{u}_{\alpha}$ at position $\mathbf{x}$ and time $t$ along the discrete velocity $\mathbf{c}_i$. $\delta t$ is the time step, $\mathbf{\Lambda}=(\mathbf{\Lambda}_{ik})$ is a $q \times q$ invertible collision matrix with $q$ representing the number of discrete velocities. $\bar{D}_i=\partial_t +\gamma \mathbf{c}_i\cdot \nabla$ with the parameter $\gamma \in \{0,1\}$, which can be discretized by some different first-order difference schemes \cite{Chai2020}. $f_{i, \alpha}^{eq}(\mathbf{x}, t)$ is the equilibrium distribution function, $G_{i, \alpha}(\mathbf{x}, t)$ is the auxiliary distribution function, $F_{i, \alpha}(\mathbf{x}, t)$ is the distribution function of the source term $\bar{F}_{\alpha}$, to derive the correct macroscopic equation (\ref{eq:2-2}), they should be defined by \cite{Chai2020}
\begin{equation}\label{eq:2-4}
f_{i, \alpha}^{eq}=\omega_{i}\big[\bar{u}_{\alpha}+\frac{\mathbf{c}_i\cdot(\bar{u}_{\alpha}\mathbf{u}+P\mathbf{E}_{\alpha})}{c_{s}^{2}}\big],
\end{equation}
\begin{equation}\label{eq:2-5}
G_{i, \alpha}=\big(1-\frac{s_{1}}{2}\big)\omega_{i}\frac{\mathbf{c}_i\cdot\partial_t(\bar{u}_{\alpha}\mathbf{u}+P\mathbf{E}_{\alpha})}{c_{s}^{2}},
\end{equation}
\begin{equation}\label{eq:2-6}
F_{i, \alpha}=\omega_{i}\bar{F}_{\alpha},
\end{equation}
where the simple linear equilibrium distribution function (\ref{eq:2-4}) with respect to the discrete velocity is considered. $\omega_{i}$ is the weight coefficient, $s_{1}$ is the relaxation parameter corresponding to the kinematic viscosity $\nu$. $c_s$ is the lattice sound speed related to lattice speed $c=\delta x/\delta t$ with $\delta x$ being the lattice spacing. In the DdQq (q discrete velocities in $d$-dimensional space) lattice model, the discrete velocity $\mathbf{c}_i$, the weight coefficient $\omega_{i}$ and the lattice sound speed $c_{s}$ should satisfy the following relations,
\begin{subequations}\label{eq:DdQq}
\begin{equation}
\sum_{i}\omega_{i}=1,
\end{equation}
\begin{equation}
\sum_{i}\omega_{i}\mathbf{c}_{i}=\mathbf{0},
\end{equation}
\begin{equation}
\sum_{i}\omega_{i}\mathbf{c}_{i}\mathbf{c}_{i}=c_{s}^{2}\mathbf{I},
\end{equation}
\end{subequations}
where $\mathbf{I}$ is the unit matrix. Here we list some special cases that have been widely used in the LBM.

\noindent
D1Q2 lattice model:
\begin{subequations}\label{eq:d1q2}
\begin{equation}
\mathbf{c}_{i}=(1, -1)c,
\end{equation}
\begin{equation}
\omega_{1}=\omega_{2}=\frac{1}{2},
\end{equation}
\begin{equation}
c_{s}=c.
\end{equation}
\end{subequations}

\noindent
D1Q3 lattice model:
\begin{subequations}\label{eq:d1q3}
\begin{equation}
\mathbf{c}_{i}=(0, 1, -1)c,
\end{equation}
\begin{equation}
\omega_{0}=\frac{2}{3}, \ \omega_{1}=\omega_{2}=\frac{1}{6},
\end{equation}
\begin{equation}
c_{s}=\frac{c}{\sqrt{3}}.
\end{equation}
\end{subequations}

\noindent
D2Q4 lattice model:
\begin{subequations}\label{eq:d2q4}
\begin{equation}
\mathbf{c}_{i}=\left(\begin{array}{cccc}
 1 & 0 & -1 & 0  \\
 0 & 1 & 0 & -1
\end{array} \right)c
\end{equation}
\begin{equation}
\omega_{1}=\omega_{2}=\omega_{3}=\omega_{4}=\frac{1}{4},
\end{equation}
\begin{equation}
c_{s}=\frac{c}{\sqrt{2}}.
\end{equation}
\end{subequations}

\noindent
D2Q5 lattice model:
\begin{subequations}\label{eq:d2q5}
\begin{equation}
\mathbf{c}_{i}=\left(\begin{array}{ccccc}
0 & 1 & 0 & -1 & 0  \\
0 & 0 & 1 & 0 & -1
\end{array} \right)c
\end{equation}
\begin{equation}
\omega_{0}=\frac{1}{3}, \ \omega_{1}=\omega_{2}=\omega_{3}=\omega_{4}=\frac{1}{6},
\end{equation}
\begin{equation}
c_{s}=\frac{c}{\sqrt{3}}.
\end{equation}
\end{subequations}

\noindent
D2Q9 lattice model:
\begin{subequations}\label{eq:d2q9}
\begin{equation}
\mathbf{c}_{i}=\left(\begin{array}{ccccccccc}
0 & 1 & 0 & -1 & 0 & 1 & -1 & -1 & 1 \\
0 & 0 & 1 & 0 & -1 & 1 & 1 &-1 & -1
\end{array} \right)c
\end{equation}
\begin{equation}
\omega_{0}=\frac{4}{9}, \ \omega_{1-4}=\frac{1}{9}, \ \omega_{5-8}=\frac{1}{36},
\end{equation}
\begin{equation}
c_{s}=\frac{c}{\sqrt{3}}.
\end{equation}
\end{subequations}

\noindent
D3Q6 lattice model:
\begin{subequations}\label{eq:d3q6}
\begin{equation}
\mathbf{c}_{i}=\left(\begin{array}{cccccc}
1 & -1 & 0 & 0 & 0 & 0 \\
0 & 0 & 1 & -1 & 0 & 0 \\
0 & 0 & 0 & 0 & 1 & -1
\end{array} \right)c
\end{equation}
\begin{equation}
\omega_{1-6}=\frac{1}{6},
\end{equation}
\begin{equation}
c_{s}=\frac{c}{\sqrt{3}}.
\end{equation}
\end{subequations}

\noindent
D3Q7 lattice model:
\begin{subequations}\label{eq:d3q7}
\begin{equation}
\mathbf{c}_{i}=\left(\begin{array}{ccccccc}
0 & 1 & -1 & 0 & 0 & 0 & 0 \\
0 & 0 & 0 & 1 & -1 & 0 & 0 \\
0 & 0 & 0 & 0 & 0 & 1 & -1
\end{array} \right)c
\end{equation}
\begin{equation}
\omega_{0}=\frac{1}{4}, \ \omega_{1-6}=\frac{1}{8},
\end{equation}
\begin{equation}
c_{s}=\frac{c}{2}.
\end{equation}
\end{subequations}

\noindent
D3Q15 lattice model:
\begin{subequations}\label{eq:d3q15}
\begin{equation}
\mathbf{c}_{i}=\left(\begin{array}{ccccccccccccccc}
0 & 1 & -1 & 0 & 0 & 0 & 0 & 1 & 1 & 1 & -1 & -1 & -1 & 1  & -1 \\
0 & 0 & 0 & 1 & -1 & 0 & 0 & 1 & 1 & -1 & 1 & -1 & 1  & -1 & -1\\
0 & 0 & 0 & 0 & 0 & 1 & -1 & 1 & -1 & 1 & 1 & 1  & -1 & -1 & -1
\end{array} \right)c
\end{equation}
\begin{equation}
\omega_{0}=\frac{2}{9}, \ \omega_{1-6}=\frac{1}{9}, \ \omega_{7-14}=\frac{1}{72}
\end{equation}
\begin{equation}
c_{s}=\frac{c}{\sqrt{3}}.
\end{equation}
\end{subequations}
We would like to point out that different lattice models can be applied for different evolution equations represented by different $\alpha$. In the MDF-LBM for the convection-diffusion system-based NSEs, the macroscopic variable $\bar{u}_{\alpha}$ is computed by
\begin{equation}\label{eq:2-7}
\bar{u}_{\alpha}=\sum_i f_{i, \alpha}.
\end{equation}

\section{The direct Taylor expansion of the multiple-distribution-function lattice Boltzmann method}

Historically, there are some asymptotic analysis methods that can be used to derive the macroscopic governing equation (\ref{eq:2-2}) from the MDF-LBM \cite{Chai2020}, including the Chapman-Enskog analysis \cite{Chen1998, Chapman1970}, Maxwell iteration method \cite{Yong2016}, direct Taylor expansion method \cite{Wagner2006}, recurrence equations method \cite{dHumieres2009}. However, it has been shown that at the second order of expansion parameters, these four analysis methods can give the same macroscopic equations \cite{Chai2020}. For this reason, we only consider the direct Taylor expansion (DTE) method for its simplicity, and additionally, compared to the commonly used Chapman-Enskog analysis, this method only includes a single expansion parameter $\delta t$.

\subsection{The direct Taylor expansion}
Based on the previous works \cite{Chai2020, Wagner2006}, when the Taylor expansion is applied to Eq. (\ref{eq:2-3}), we have
\begin{equation}\label{eq:3-1}
\sum_{j=1}^{N}\frac{\delta t^j}{j!}D_i^j f_{i, \alpha} +O(\delta t^{N+1})=-\mathbf{\Lambda}_{ik} (f_{k, \alpha}-f_{k, \alpha}^{eq})+\delta t \big[G_{i, \alpha}(\mathbf{x},t)+F_{i, \alpha}(\mathbf{x}, t)+\frac{\delta t}{2}\bar{D}_i F_{i, \alpha}(\mathbf{x}, t)\big],
\end{equation}
where $D_{i}=\partial_t+\mathbf{c}_{i}\cdot\nabla$. Introduce $f_{i, \alpha}^{ne}=f_{i, \alpha}-f_{i, \alpha}^{eq}$ and substitute it into the collision term of Eq. (\ref{eq:3-1}), we can derive the following equations,
\begin{subequations}\label{eq:3-2}
\begin{equation}
f_{i, \alpha}^{ne}=O(\delta t),
\end{equation}
\begin{equation}
\sum_{j=1}^{N-1}\frac{\delta t^j}{j!}D_i^j (f_{i, \alpha}^{eq}+f_{i, \alpha}^{ne}) + \frac{\delta t^N}{N!}D_i^N f_{i, \alpha}^{eq}=-\mathbf{\Lambda}_{ik} f_{k, \alpha}^{ne}+\delta t \big[G_{i, \alpha}(\mathbf{x},t)+F_{i, \alpha}(\mathbf{x}, t)+\frac{\delta t}{2}\bar{D}_i F_{i, \alpha}(\mathbf{x}, t)\big]+O(\delta t^{N+1}).
\end{equation}
\end{subequations}
Then the equations at first and second orders of $\delta t$ can be obtained,
\begin{subequations}\label{eq:3-3}
\begin{equation}
D_i f_{i, \alpha}^{eq}=-\frac{1}{\delta t }\mathbf{\Lambda}_{ik} f_{k, \alpha}^{ne}+(G_{i, \alpha}+F_{i, \alpha})+O(\delta t),
\end{equation}
\begin{equation}
D_i (f_{i, \alpha}^{eq}+f_{i, \alpha}^{ne})+\frac{\delta t}{2}D_i^2 f_{i, \alpha}^{eq}=-\frac{1}{\delta t }\mathbf{\Lambda}_{ik} f_{k, \alpha}^{ne}+\big(G_{i, \alpha}+F_{i, \alpha}+\frac{\delta t}{2}\bar{D}_{i}F_{i, \alpha}\big)+O(\delta t^2).
\end{equation}
\end{subequations}
From Eq. (\ref{eq:3-3}a), we have
\begin{equation}\label{eq:3-4}
\frac{\delta t}{2}D_i^2 f_{i, \alpha}^{eq}=-\frac{1}{2} D_i \mathbf{\Lambda}_{ik} f_{k, \alpha}^{ne}+\frac{\delta t}{2}D_i(G_{i, \alpha}+F_{i, \alpha})+O(\delta t^2).
\end{equation}
Substituting Eq. (\ref{eq:3-4}) into Eq. (\ref{eq:3-3}b) yields
\begin{equation}\label{eq:3-5}
D_i f_{i, \alpha}^{eq}+D_i\big(\delta_{ik}-\frac{1}{2}\mathbf{\Lambda}_{ik}\big)f_{k, \alpha}^{ne}+\frac{\delta t}{2}D_i(G_{i, \alpha}+F_{i, \alpha})\nonumber\\=-\frac{1}{\delta t}\mathbf{\Lambda}_{ik} f_{k, \alpha}^{ne}+G_{i, \alpha}+F_{i, \alpha}+\frac{\delta t}{2}\bar{D}_iF_{i, \alpha}+O(\delta t^2).
\end{equation}

As pointed out in the previous work \cite{Chai2020}, to recover Eq. (\ref{eq:2-2}) from the MDF-LBM (\ref{eq:2-3}), some appropriate requirements on the collision matrix $\mathbf{\Lambda}$ are needed,
\begin{eqnarray}\label{eq:3-6}
\sum_i \mathbf{e}_{i}\mathbf{\Lambda}_{ik} = s_0 \mathbf{e}_{k},\ \ \sum_i \mathbf{c}_{i}\mathbf{\Lambda}_{ik} = s_1 \mathbf{c}_k,
\end{eqnarray}
where $\mathbf{e}=(1,1,\cdots,1)\in R^q$, $s_0$ and $s_1$ are eigenvalues of the collision matrix $\mathbf{\Lambda}$ or the relaxation parameters corresponding to the zero- and first-order moments of the distribution function. Here it should be noted that a more general case of Eq. (\ref{eq:3-6}) shown in Ref. \cite{Chai2020} can also be considered.

According to Eqs. (\ref{eq:2-4}), (\ref{eq:2-5}) and (\ref{eq:2-6}), one can also determine the moments of $f_{i, \alpha}^{eq}(\mathbf{x},t)$, $G_{i, \alpha, }(\mathbf{x},t)$ and $F_{i, \alpha}(\mathbf{x},t)$,
\begin{subequations}\label{eq:3-7}
\begin{equation}
\sum_i f_{i, \alpha}^{eq}=\bar{u}_{\alpha},\ \ \sum_i \mathbf{c}_i f_{i, \alpha}^{eq}=\bar{u}_{\alpha}\mathbf{u}+P\mathbf{E}_{\alpha},\ \ \sum_i \mathbf{c}_i \mathbf{c}_i
f_{i, \alpha}^{eq}=\bar{u}_{\alpha}c_s^2 \mathbf{I},
\end{equation}
\begin{equation}
\sum_i G_{i, \alpha}=0,\ \ \sum_i \mathbf{c}_i G_{i, \alpha}=\big(1-\frac{s_{1}}{2}\big)\partial_t(\bar{u}_{\alpha}\mathbf{u}+P\mathbf{E}_{\alpha}),
\end{equation}
\begin{equation}
\sum_i F_{i, \alpha}=\bar{F}_{\alpha},\ \ \sum_i \mathbf{c}_i F_{i, \alpha}=\mathbf{0}.
\end{equation}
\end{subequations}
From Eqs. (\ref{eq:2-7}) and (\ref{eq:3-7}a), we can first obtain
\begin{equation}\label{eq:3-8}
\sum_i f_{i, \alpha}^{ne}=0.
\end{equation}
Then with the help of Eq. (\ref{eq:3-6}), one can also derive the following equations,
\begin{equation}\label{eq:3-9}
\sum_i \mathbf{e}_{i}\mathbf{\Lambda}_{ik}f_{k, \alpha}^{ne} = s_0 \sum_k\mathbf{e}_{k}f_{k, \alpha}^{ne} = \mathbf{0},
\end{equation}
\begin{equation}\label{eq:3-10}
\sum_i \mathbf{c}_{i}\mathbf{\Lambda}_{ik}f_{k, \alpha}^{ne} = s_1 \sum_k\mathbf{c}_k f_{k, \alpha}^{ne}.
\end{equation}

\subsection{Derivation of the convection-diffusion-system based Navier-Stokes-equations}

We now present some details on how to derive the convection-diffusion-system based NSEs (\ref{eq:2-2}) from MDF-LBM (\ref{eq:2-3}). To this end, we conduct a summation of Eq. (\ref{eq:3-5}), and obtain the following equation,
\begin{equation}\label{eq:3-11}
 \frac{\partial\bar{u}_\alpha}{\partial t}+\nabla\cdot(\bar{u}_\alpha\mathbf{u}+P\mathbf{E}_\alpha)+\nabla\cdot\big[\big(1-\frac{s_1}{2}\big)\sum_i \mathbf{c}_{i}f_{i, \alpha}^{ne}\big]+\frac{\delta t}{2}\nabla\cdot\big[\big(1-\frac{s_{1}}{2}\big)\partial_t(\bar{u}_{\alpha}\mathbf{u}+P\mathbf{E}_{\alpha})\big]\nonumber\\=\bar{F}_\alpha+O(\delta t^2).
\end{equation}
Now let us give an evaluation to the term $\sum_i \mathbf{c}_{i}f_{i, \alpha}^{ne}$. Actually, from Eq. (\ref{eq:3-3}a) we have
\begin{eqnarray}\label{eq:3-12}
\sum_{i} \mathbf{c}_{i}f_{i, \alpha}^{ne} & = & -\delta t\sum_{i} \mathbf{c}_{i}\mathbf{\Lambda}_{ik}^{-1}\big(D_k f_{k, \alpha}^{eq}-G_{k, \alpha}-F_{k, \alpha}\big)+O(\delta t^{2})\nonumber \\
& = &-\frac{\delta t}{s_{1}}\sum_{i} \mathbf{c}_{i}\big(D_i f_{i, \alpha}^{eq}-G_{i, \alpha}-F_{i, \alpha}\big)+O(\delta t^{2})\nonumber \\
& = &-\frac{\delta t}{2}\partial_t(\bar{u}_{\alpha}\mathbf{u}+P\mathbf{E}_{\alpha})-\frac{\delta t}{s_{1}}c_{s}^{2}\nabla\bar{u}_{\alpha}+O(\delta t^{2}),
\end{eqnarray}
where Eqs. (\ref{eq:3-6}) and (\ref{eq:3-7}) have been used.
Substituting Eq. (\ref{eq:3-12}) into Eq. (\ref{eq:3-11}) yields
\begin{equation}\label{eq:3-13}
 \frac{\partial\bar{u}_\alpha}{\partial t}+\nabla\cdot(\bar{u}_\alpha\mathbf{u}+P\mathbf{E}_\alpha)=\nabla\cdot\big[\big(\frac{1}{s_{1}}-\frac{1}{2}\big)c_{s}^{2}\delta t\nabla\bar{u}_{\alpha}\big]+\bar{F}_\alpha+O(\delta t^2).
\end{equation}
If we neglect the truncation error $O(\delta t^2)$, one can obtain the macroscopic convection-diffusion-system based NSEs (\ref{eq:2-2}) with the following viscosity,
\begin{equation} \label{eq:3-13a}
\nu=\big(\frac{1}{s_{1}}-\frac{1}{2}\big)c_{s}^{2}\delta t.
\end{equation}
Here we would also like to give a special discussion on how to calculate the pressure $P$. From Eq. (\ref{eq:3-12}) we can get
\begin{eqnarray}\label{eq:3-14}
\sum_{i} \mathbf{c}_{i,\alpha}f_{i, \alpha}^{ne} & = & \sum_{i} \mathbf{c}_{i,\alpha}f_{i, \alpha}-\sum_{i} \mathbf{c}_{i,\alpha}f_{i, \alpha}^{eq} \nonumber \\
& = & \sum_{i} \mathbf{c}_{i,\alpha}f_{i, \alpha}-(\bar{u}_{\alpha}u_{\alpha}+P) \nonumber \\
& = &-\frac{\delta t}{s_{1}}c_{s}^{2}\nabla_{\alpha}\bar{u}_{\alpha}+O(\delta t^{2}+\delta t \mbox{Ma}^{2}), \ (\alpha\neq 0),
\end{eqnarray}
where Ma is the Mach number.
Neglecting the truncation error term $O(\delta t^{2}+\delta t \mbox{Ma}^{2})$ and summing Eq. (\ref{eq:3-14}) over $\alpha$, one can derive
\begin{eqnarray}\label{eq:3-15a}
\sum_{\alpha=1}^{d}\sum_{i} \mathbf{c}_{i,\alpha}f_{i, \alpha}-(|\mathbf{u}|^{2}+d P) =-\frac{\delta t}{s_{1}}c_{s}^{2}R,
\end{eqnarray}
where Eq. (\ref{eq:2-1}a) has been used.
From Eq. (\ref{eq:3-15a}) we can give an expression to compute the pressure,
\begin{equation}\label{eq:3-15}
P=\frac{1}{d}\big(\sum_{\alpha=1}^{d}\sum_{i} \mathbf{c}_{i,\alpha}f_{i, \alpha}+\frac{\delta t}{s_{1}}c_{s}^{2}R-|\mathbf{u}|^{2}\big).
\end{equation}
From above procedure, it can be found that the continuity equation (\ref{eq:2-1}a) has been used to derive Eq. (\ref{eq:3-15}), this means that the formula for the pressure [Eq. (\ref{eq:3-15})] is consistent with the continuity equation. In other words, we do not need to consider the continuity equation (\ref{eq:2-1}a) in the present MDF-LBM, and the evolution equation (\ref{eq:2-3}) with $\alpha=0$ can be omitted. We also note that the term related to $G_{i, \alpha}$ in Eq. (\ref{eq:2-3}) can also be neglected since it is order of $O(\delta t\mbox{Ma}^{2})$. In this case, the evolution equation (\ref{eq:2-3}) can be simplified by
\begin{equation}\label{eq:2-3a}
 f_{i, \alpha}(\mathbf{x}+\mathbf{c}_i \delta t,t+\delta t)=f_{i, \alpha}(\mathbf{x}, t)-\mathbf{\Lambda}_{ik} \big[f_{k, \alpha}(\mathbf{x}, t)-f_{k, \alpha}^{eq}(\mathbf{x}, t)\big]+\delta t \big[F_{i, \alpha}(\mathbf{x}, t)+\frac{\delta t}{2}\bar{D}_i F_{i, \alpha}(\mathbf{x}, t)\big].
\end{equation}
Then one can derive two special schemes \cite{Chai2016a} from Eq. (\ref{eq:2-3a}). \\
\textbf{Scheme I}: $\gamma=0$. Under this condition, $\bar{D}_i F_{i, \alpha}(\mathbf{x}, t)=\partial_t F_{i, \alpha}(\mathbf{x}, t)$. Considering the Euler method for the time derivative, we have
\begin{equation}\label{eq:2-3b}
 f_{i, \alpha}(\mathbf{x}+\mathbf{c}_i \delta t,t+\delta t)=f_{i, \alpha}(\mathbf{x}, t)-\mathbf{\Lambda}_{ik} \big[f_{k, \alpha}(\mathbf{x}, t)-f_{k, \alpha}^{eq}(\mathbf{x}, t)\big]+\frac{\delta t}{2} \big[3F_{i, \alpha}(\mathbf{x}, t)-F_{i, \alpha}(\mathbf{x}, t-\delta t)\big].
\end{equation}
\textbf{Scheme II}: $\gamma=1$. With this choice, we have $\bar{D}_i F_{i, \alpha}(\mathbf{x}, t)=D_i F_{i, \alpha}(\mathbf{x}, t)$, which can be discretized by the following implicit finite-difference scheme,
\begin{equation}
D_i F_{i, \alpha}(\mathbf{x}, t)=\frac{F_{i, \alpha}(\mathbf{x}+\mathbf{c}_{i}\delta t, t+\delta t)-F_{i, \alpha}(\mathbf{x}, t)}{\delta t}.
\end{equation}
Substituting above equation into Eq. (\ref{eq:2-3a}) and introducing a new variable $\bar{f}_{i, \alpha}(\mathbf{x}, t)=f_{i, \alpha}(\mathbf{x}, t)-\delta t F_{i, \alpha}(\mathbf{x}, t)/2$, one can obtain
\begin{equation}\label{eq:2-3c}
 \bar{f}_{i, \alpha}(\mathbf{x}+\mathbf{c}_i \delta t,t+\delta t)=\bar{f}_{i, \alpha}(\mathbf{x}, t)-\mathbf{\Lambda}_{ik} \big[\bar{f}_{k, \alpha}(\mathbf{x}, t)-f_{k, \alpha}^{eq}(\mathbf{x}, t)\big]+\delta t \big(\delta_{ik}-\frac{1}{2}\mathbf{\Lambda}_{ik} \big)F_{k, \alpha}(\mathbf{x}, t).
\end{equation}
In this scheme, the macroscopic variable $\bar{u}_{\alpha}$ is computed by
\begin{equation}\label{eq:2-7a}
\bar{u}_{\alpha}=\sum_i f_{i, \alpha}=\sum_i \bar{f}_{i, \alpha}+\frac{\delta t}{2}\bar{F}_{\alpha},
\end{equation}
while the pressure is still given by Eq. (\ref{eq:3-15}) in which $f_{i, \alpha}$ is replaced by $\bar{f}_{i, \alpha}$.

\subsection{The computational schemes for the velocity gradient, velocity divergence, strain rate tensor, shear stress and vorticity}

Besides the fluid velocity and pressure mentioned above, usually we also need to consider some other physical variables. For instance, the strain rate tensor (or the symmetric velocity gradient tensor), shear stress and vorticity (it is related to the antisymmetric velocity gradient tensor) are also important in the study of the non-Newtonian fluid flows and turbulence \cite{Bird1987, McComb1990}, and in the framework of LBM, they also received increasing attention in the past years \cite{Artoli2004, Kruger2009, Chai2012, Yong2012, Peng2017, Hajabdollahi2020}. In the SRT-LB model, Artoli et al. \cite{Artoli2004} first developed a local scheme for the shear stress where the second-order moment of non-equilibrium distribution function is adopted, and applied the scheme to study the blood flows in a symmetric bifurcation. Then Kr\"{u}ger et al. \cite{Kruger2009} conducted a theoretical analysis, and found that the the local scheme for the shear stress has a second-order convergence rate. Chai and Zhao \cite{Chai2012} further considered the forcing term effect on the computation of the strain rate tensor and shear stress, and proposed two more general local schemes for the strain rate tensor and shear stress. They also performed the theoretical analysis and numerical simulations, and demonstrated that the local schemes have a second-order accuracy in space. Yong and Luo \cite{Yong2012} carried out an asymptotic analysis on the SRT-LB model coupled with the simple bounce-back boundary condition, and illustrated that the shear stress computed with the second-order moment of the non-equilibrium distribution function has a second-order accuracy in space. In addition, the velocity divergence can also be calculated locally by the second-order moment of non-equilibrium distribution function \cite{Chai2012}.

In the LB method, it seems difficult to construct the local schemes for the velocity gradient and vorticity, and for this reason, some non-local finite-difference schemes are usually adopted (e.g., Ref. \cite{Luo2011}). Recently, through the careful design of the high-order moments of equilibrium distribution function, Peng et al. \cite{Peng2017} developed a local scheme for the vorticity in the MRT-LB model with D3Q27 lattice structure. However, this scheme cannot be extended to other lattice models, e.g., the commonly used D2Q9 lattice model in two-dimensional space. To overcome this limitation, Hajabdollahi and Premnath \cite{Hajabdollahi2020} proposed another scheme for the computation of vorticity in the double-distrbution-function LBM. However, besides the LB model for NSEs, they also need to introduce another LB model for an additional CDE, which would bring more computational cost. In this work, we will develop a local scheme for the velocity gradient, which can be further extended to calculate the velocity divergence, the strain rate tensor, shear stress and vorcitity without introducing any additional requirements. Compared to the previous works \cite{Artoli2004, Kruger2009, Chai2012, Yong2012, Peng2017, Hajabdollahi2020} where second-order moments of the non-equilibrium distribution function are adopted, the present local schemes for velocity gradient, velocity divergence, the strain rate tensor, shear stress and vorticity only include the first-order moments of the non-equilibrium distribution function, and what is more, they are not restricted to the special lattice models.

Now let focus on how to calculate the the velocity gradient, velocity divergence, strain rate tensor shear stress and vorticity in the MDF-LBM. Actually, from Eq. (\ref{eq:3-12}) we can first obtain the velocity gradient,
\begin{eqnarray}\label{eq:3-16}
\nabla_{\beta}u_{\alpha}=-\frac{s_{1}}{ c_{s}^{2}\delta t}\sum_{i} \mathbf{c}_{i,\beta}f_{i, \alpha}^{ne}, \ (\alpha\neq0),
\end{eqnarray}
where the truncation error in Eq. (\ref{eq:3-14}) has been neglected. We note that Eq. (\ref{eq:3-16}) is similar to the schemes reported in the previous works \cite{Chai2013, Chai2014}, and can be used to derive the schemes for the velocity divergence $\nabla\cdot \mathbf{u}$, strain rate tensor $\mathbf{S}$, shear stress $\sigma$ and the antisymmetric velocity gradient tensor $\mathbf{\Omega}$,
\begin{eqnarray}\label{eq:3-16a}
\nabla\cdot \mathbf{u}=\sum_{\alpha=1}^{d}\nabla_{\alpha}u_{\alpha}=-\frac{s_{1}}{ c_{s}^{2}\delta t}\sum_{\alpha=1}^{d}\sum_{i} \mathbf{c}_{i,\alpha}f_{i, \alpha}^{ne},
\end{eqnarray}
\begin{eqnarray}\label{eq:3-17}
S_{\alpha\beta}=\frac{1}{2}(\nabla_{\beta}u_{\alpha}+\nabla_{\alpha}u_{\beta})=-\frac{s_{1}}{2c_{s}^{2}\delta t}\sum_{i} (\mathbf{c}_{i,\beta}f_{i, \alpha}^{ne}+\mathbf{c}_{i,\alpha}f_{i, \beta}^{ne}), \ (\alpha\neq0, \beta\neq0),
\end{eqnarray}
\begin{eqnarray}\label{eq:3-17a}
\sigma_{\alpha\beta}=2\rho_{0}\nu S_{\alpha\beta}=-\rho_{0}\big(1-\frac{s_{1}}{2}\big)\sum_{i} (\mathbf{c}_{i,\beta}f_{i, \alpha}^{ne}+\mathbf{c}_{i,\alpha}f_{i, \beta}^{ne}), \ (\alpha\neq0, \beta\neq0),
\end{eqnarray}
\begin{eqnarray}\label{eq:3-18}
\Omega_{\alpha\beta}=\frac{1}{2}(\nabla_{\beta}u_{\alpha}-\nabla_{\alpha}u_{\beta})=-\frac{s_{1}}{2c_{s}^{2}\delta t}\sum_{i} (\mathbf{c}_{i,\beta}f_{i, \alpha}^{ne}-\mathbf{c}_{i,\alpha}f_{i, \beta}^{ne}), \ (\alpha\neq0, \beta\neq0).
\end{eqnarray}
Then based on the following relation between the antisymmetric velocity gradient tensor $\mathbf{\Omega}$ and vorticity $\mathbf{\omega}$=$\nabla\times\mathbf{u}$,
\begin{equation}\label{eq:3-19}
\Omega_{\alpha\beta}=-\frac{1}{2}\epsilon_{\alpha\beta\gamma}\omega_{\gamma},
\end{equation}
one can determine the component of vorticity $\omega_{\gamma}$, $\epsilon_{\alpha\beta\gamma}$ is the Levi-Civita tensor.

Finally, we also give some remarks on the present MDF-LBM for the convection-diffusion-system based incompressible NSEs. \\
\textbf{Remark I}: In the above analysis, if the diffusive scaling ($\delta t\propto\delta x^{2}$) is considered \cite{Zhang2019,Junk2005}, the evolution equation (\ref{eq:2-3}) can be simply written as
\begin{equation}\label{eq:3-21}
 f_{i, \alpha}(\mathbf{x}+\mathbf{c}_i \delta t,t+\delta t)=f_{i, \alpha}(\mathbf{x}, t)-\mathbf{\Lambda}_{ik} \big[f_{k, \alpha}(\mathbf{x}, t)-f_{k, \alpha}^{eq}(\mathbf{x}, t)\big]+\delta t F_{i, \alpha}(\mathbf{x}, t).
\end{equation}
Then we can also derive the macroscopic equation (\ref{eq:3-13}) from Eq. (\ref{eq:3-21}), while the truncation error is $O(\delta x^{2})$ rather than $O(\delta t^{2})$. On the other hand, for the specified kinematic viscosity $\nu$ and the relaxation parameter $s_{1}$, one can also obtain the diffusive scaling $\delta t\propto\delta x^{2}$ from Eq. (\ref{eq:3-13a}), which means that the diffusive scaling used in the LBM is reasonable. Additionally, the computational schemes for the velocity, pressure, velocity gradient, velocity divergence, strain rate tensor, shear stress and vorticity are the same as Eqs. (\ref{eq:2-7}), (\ref{eq:3-15}), (\ref{eq:3-16}), (\ref{eq:3-16a}), (\ref{eq:3-17}), (\ref{eq:3-17a}) and (\ref{eq:3-18}). In the following, for simplicity we would consider the MDF-LBM (\ref{eq:3-21}) for the convection-diffusion-system based incompressible NSEs.\\
\textbf{Remark II}: In the single LBM for the NSEs \cite{Artoli2004, Kruger2009, Chai2012, Yong2012}, the velocity divergence, strain rate tensor and shear stress can be computed locally with the second-order moments of the non-equilibrium distribution function, while in the present MDF-LBM for the convection-diffusion-system based incompressible NSEs, only the first-order moments of the non-equilibrium distribution function are needed, as seen from Eqs. (\ref{eq:3-16a}), (\ref{eq:3-17}) and (\ref{eq:3-17a}).\\
\textbf{Remark III}: In the commonly used LBM for NSEs, it is difficult to compute the velocity gradient locally. However, in the present MDF-LBM for the convection-diffusion-system based incompressible NSEs, the velocity gradient can be calculated locally from Eq. (\ref{eq:3-16}), which can also be used to determine some other physical variables, including the velocity divergence $\nabla\cdot\mathbf{u}$, strain rate tensor $\mathbf{S}$, shear stress $\sigma$ and vorticity $\omega$. Additionally, we also note that although some local schemes have been developed for vorticity, there are still some limitations in these available works \cite{Peng2017, Hajabdollahi2020}. For example, the local scheme proposed by Peng et al. \cite{Peng2017} is only suitable for the D3Q27 lattice model, and cannot be extended to some other lattice models. Compared to the scheme in Ref. \cite{Peng2017}, the scheme developed by Hajabdollahi and Premnath \cite{Hajabdollahi2020} is more general, while an additional CDE must be introduced and solved by another LB model, which would make the computational cost more expensive. \\
\textbf{Remark IV}: In the single LBM for the NSEs \cite{Qian1992}, the pressure is related to density through the relation $P=\rho c_{s}^{2}$ where the density is calculated with the zeroth-order of distribution function, while in the LBM for incompressible NSEs \cite{Guo2000}, the pressure is determined by the zeroth-order moment of distribution function without considering the one at zeroth direction. In the present MDF-LBM, the pressure is computed by the first-order moment of distribution function, but the one with $\alpha=0$ is not included. In addition, it should be noted that the formula for the pressure [see Eq. (\ref{eq:3-15})] is consistent with the continuity equation (\ref{eq:2-1}a), and thus we do not need to consider the evolution equation (\ref{eq:2-3}) with $\alpha=0$.

\section{Numerical results and discussion}

For simplicity but without loss of generality, we only considered two-dimensional problems in this section, and adopted some two-dimensional benchmark problems, including the Poiseuille flow, the simplified four-roll mill problem and lid-driven cavity flow, to test the developed MDF-LBM with D2Q5 lattice model. In this lattice model, the collision matrix $\mathbf{\Lambda}$ is taken as
\begin{equation}\label{eq:4-1}
\mathbf{\Lambda}=\mathbf{M}^{-1}\mathbf{S}_{d}\mathbf{M}.
\end{equation}
The transformation matrix $\mathbf{M}$ and relaxation matrix $\mathbf{S}_{d}$ appeared in above equation are defined as \cite{Cui2016}
\begin{subequations}
\begin{equation}\label{eq:4-2}
\mathbf{M} = \left( {\begin{array}{*{20}{c}}
   1 & 1 & 1 & 1 & 1  \\
   0 & c & 0 & { - c} & 0  \\
   0 & 0 & c & 0 & { - c} \\
   0 & c^{2} & { - c^{2}} & c^{2} & { - c^{2}}  \\
   {-4c^{2}} & c^{2} & c^{2} & c^{2} & c^{2} \\
\end{array}} \right),
\end{equation}
\begin{equation}\label{eq:4-3}
\mathbf{S}_{d}=\mathbf{diag}(s_{0}, s_{1}, s_{1}, s_{2}, s_{2}),
\end{equation}
\end{subequations}
where $s_{i}$ is the relaxation parameter corresponding to the $i$th-order moment of the distribution function. Additionally, according to Eq. (\ref{eq:DdQq}), the relations among the weigh coefficients and lattice speed in the D2Q5 lattice model can be generally expressed as
\begin{equation}
\omega_{i}=\frac{1-\omega_{0}}{4},\  2\omega_{i}c^{2}=c_{s}^{2},\ i=1-4,
\end{equation}
where $\omega_{0}$ is considered as a free parameter within the range $(0, 1)$. Actually, if $\omega_{0}=1/5$, one can obtain $\omega_{i}=1/5\ (i=1-4)$ and $c_{s}^{2}=2c^{2}/5$. However, if $\omega_{0}=1/3$, we have $\omega_{i}=1/6\ (i=1-4)$ and $c_{s}^{2}=c^{2}/3$, which is the same as Eq. (\ref{eq:d2q5}b) and would be used in the following simulations. Unless otherwise stated, the initialization of the distribution functions is realized by their equilibrium distribution functions where the pressure and velocity are given by $P=1$ and $u_{1}=u_{2}=0$. The anti-halfway bounce-back scheme for the MDF-LBM \cite{Zhao2020b} is used to treat Dirichlet boundary conditions of velocity.

In our simulations, to test the accuracies of the MDF-LBM and the local schemes for the physical variables, the following $L^{2}$ norm of relative error ($E_{L^2}$) is adopted,
\begin{equation}\label{eq:4-2}
E_{L^2} (\phi)=\frac{\|\phi_{a}(x, y)-\phi_{n}(x, y)\|_{2}}{\|\phi_{a}(x, y)\|_{2}},
\end{equation}
where $\phi$ denotes velocity, velocity gradient, velocity divergence, strain rate tensor or vorticity, the subscripts $a$ and $n$ represent the analytical and numerical solutions of $\phi$.

\subsection{The two-dimensional Poiseuille flow}
\begin{figure}
\centering
\includegraphics[width=0.8\textwidth]{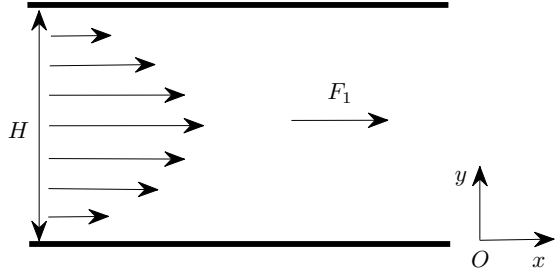}
\caption{The schematic of the Poiseuille flow.}
\label{fig1}
\end{figure}

The first problem we considered is the simple Poiseuille flow, which is driven by an external constant force in $x$ direction ($F_{1}=1.0\times 10^{-6}$), as shown in Fig. \ref{fig1}. For this problem, one can obtain its analytical solutions of velocity $\mathbf{u}=(u_{1}, u_{2})^{\top}$, velocity gradient $\nabla \mathbf{u}$, velocity divergence $\nabla\cdot \mathbf{u}$, strain rate tensor $\mathbf{S}$ and vorticity $\omega$,
\begin{subequations}\label{eq:4-1}
\begin{equation}
u_{1}=\frac{F_{1}}{2\nu}H^{2}\big[\frac{y}{H}-\big(\frac{y}{H}\big)^{2}\big], \ u_{2}=0,
\end{equation}
\begin{equation}
\frac{\partial u_{1}}{\partial x}=\frac{\partial u_{2}}{\partial x}=\frac{\partial u_{2}}{\partial y}=0,\ \frac{\partial u_{1}}{\partial y}=\frac{F_{1}}{2\nu}H\big(1-2\frac{y}{H}\big),
\end{equation}
\begin{equation}
\nabla \cdot \mathbf{u}=\frac{\partial u_{1}}{\partial x}+\frac{\partial u_{2}}{\partial y}=0,
\end{equation}
\begin{equation}
S_{xx}=S_{yy}=0,\ S_{xy}=S_{yx}=\frac{F_{1}}{4\nu}H\big(1-2\frac{y}{H}\big),
\end{equation}
\begin{equation}
\omega=\frac{\partial u_{2}}{\partial x}-\frac{\partial u_{1}}{\partial y}=-\frac{F_{1}}{2\nu}H\big(1-2\frac{y}{H}\big).
\end{equation}
\end{subequations}

\begin{figure}
\centering
\includegraphics[width=0.6\textwidth]{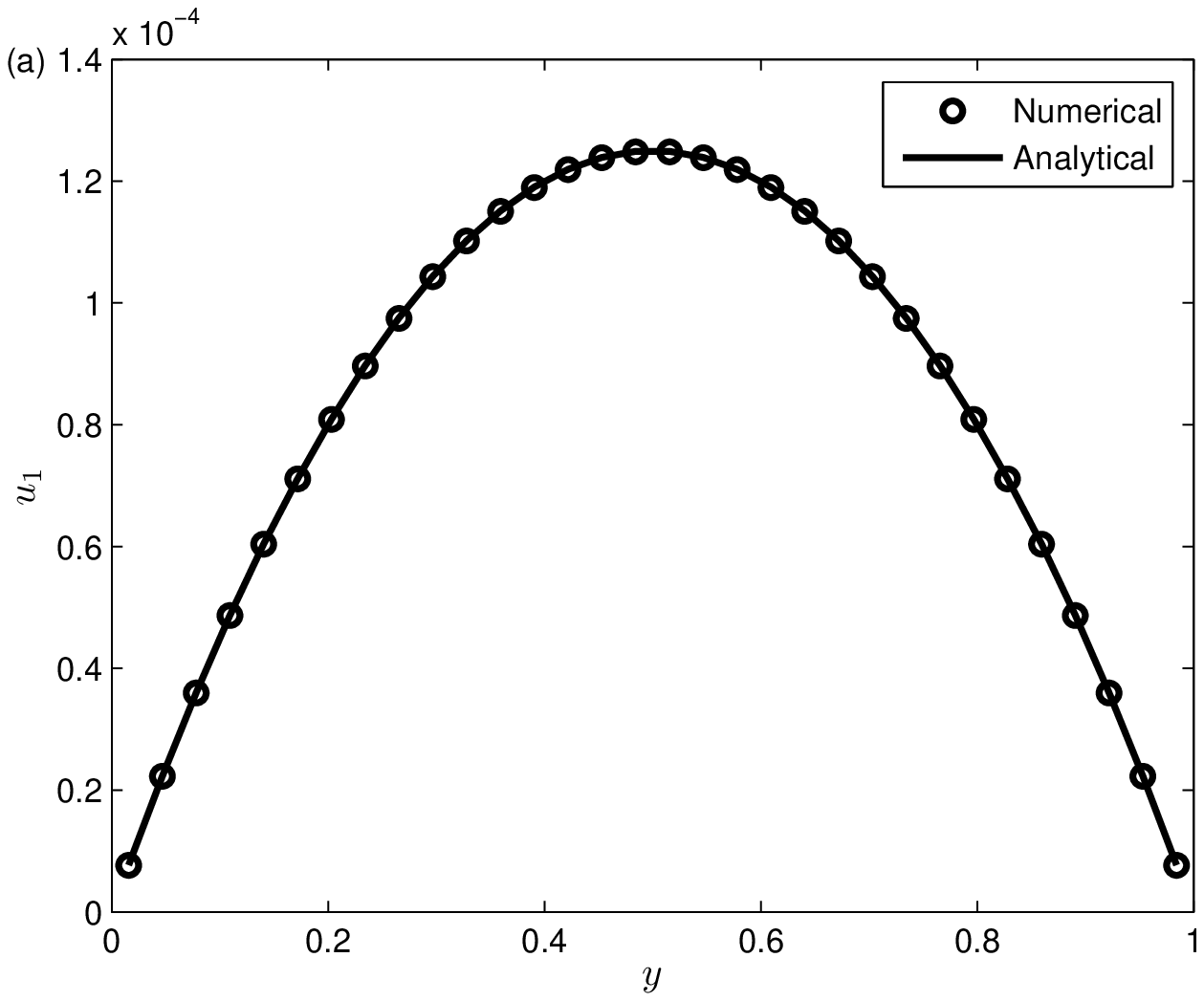}
\includegraphics[width=0.6\textwidth]{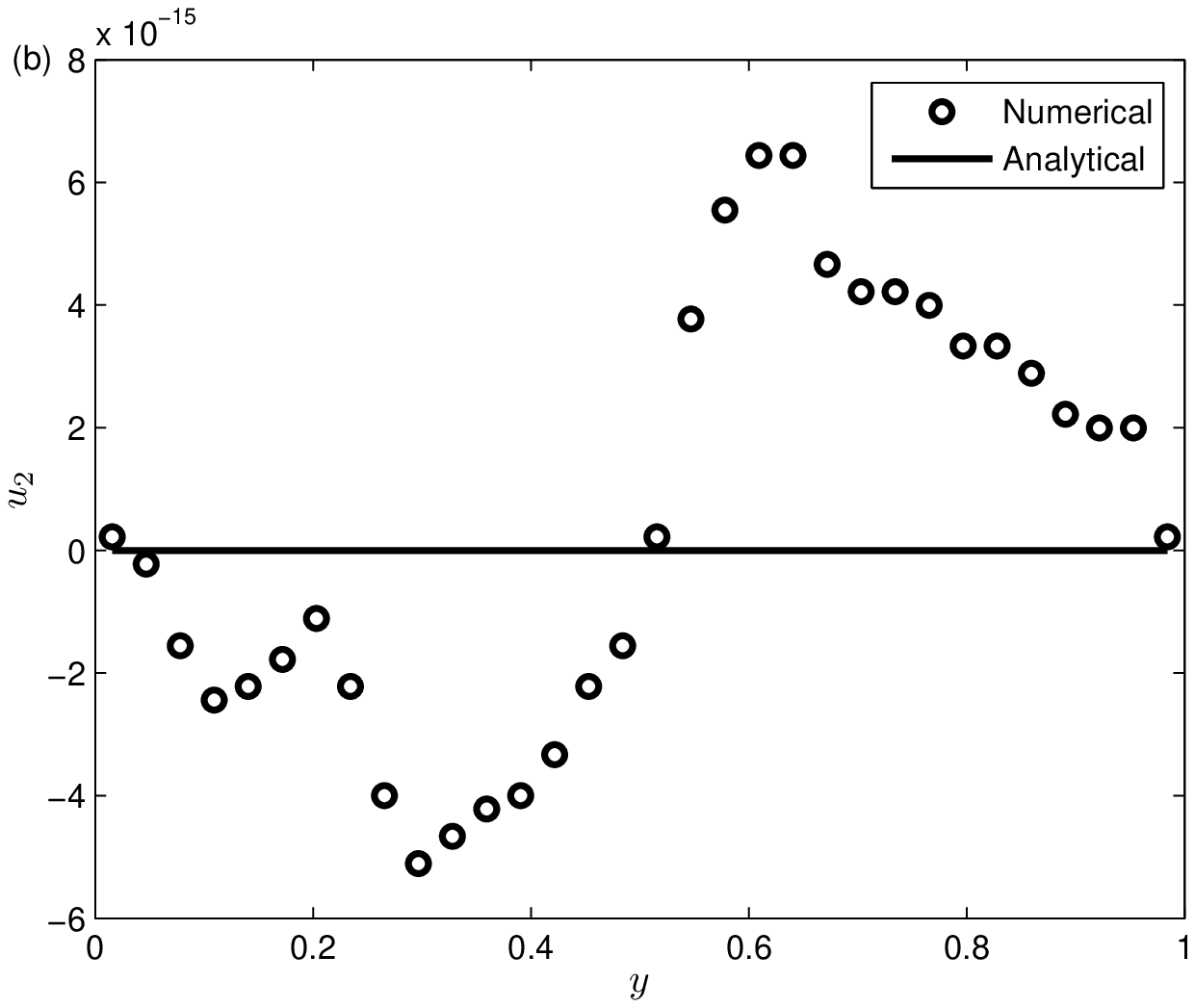}
\caption{The numerical and analytical solutions of the velocity [(a): $u_{1}$, (b): $u_{2}$].}
\label{fig2}
\end{figure}

\begin{figure}
\centering
\includegraphics[width=0.6\textwidth]{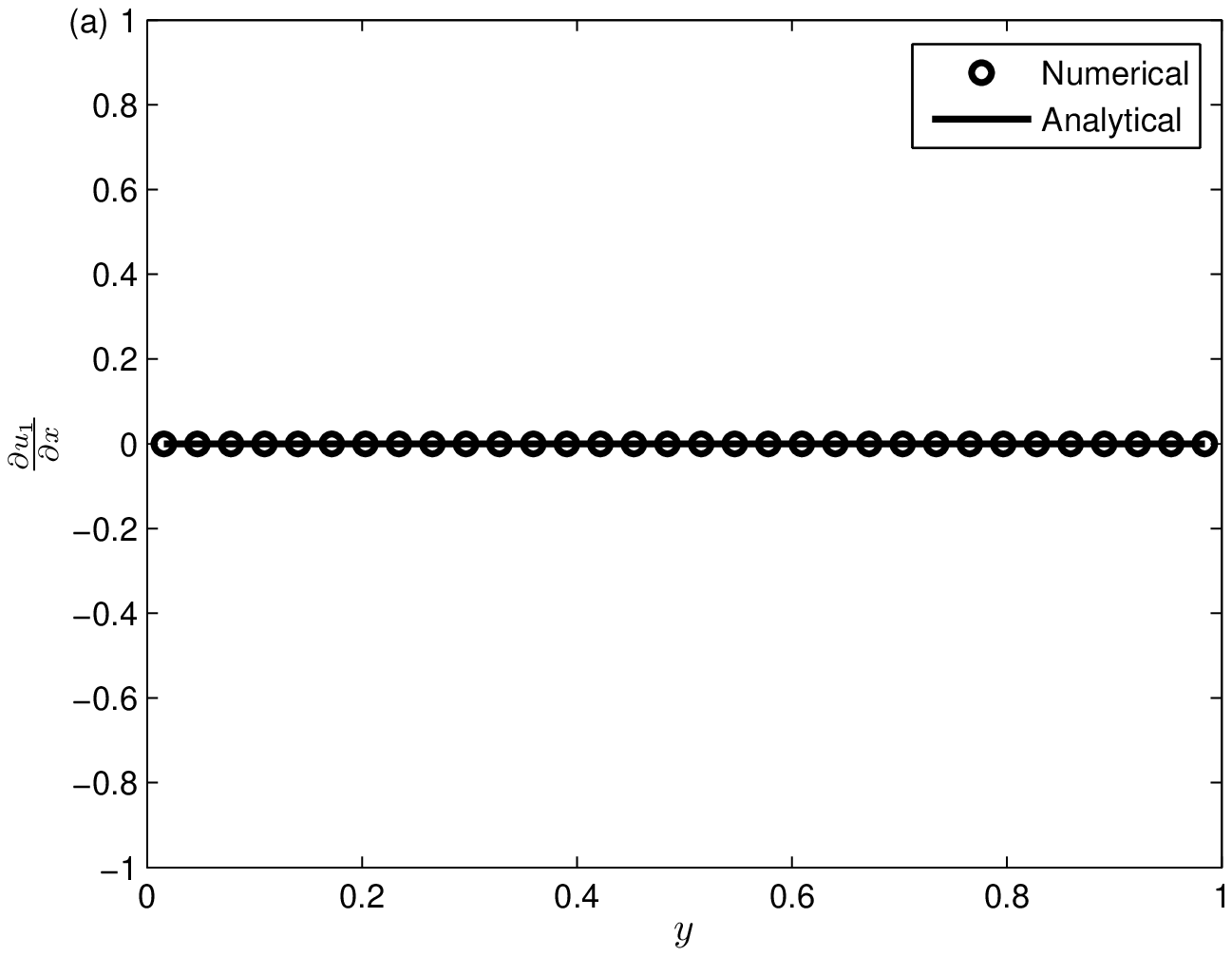}
\includegraphics[width=0.6\textwidth]{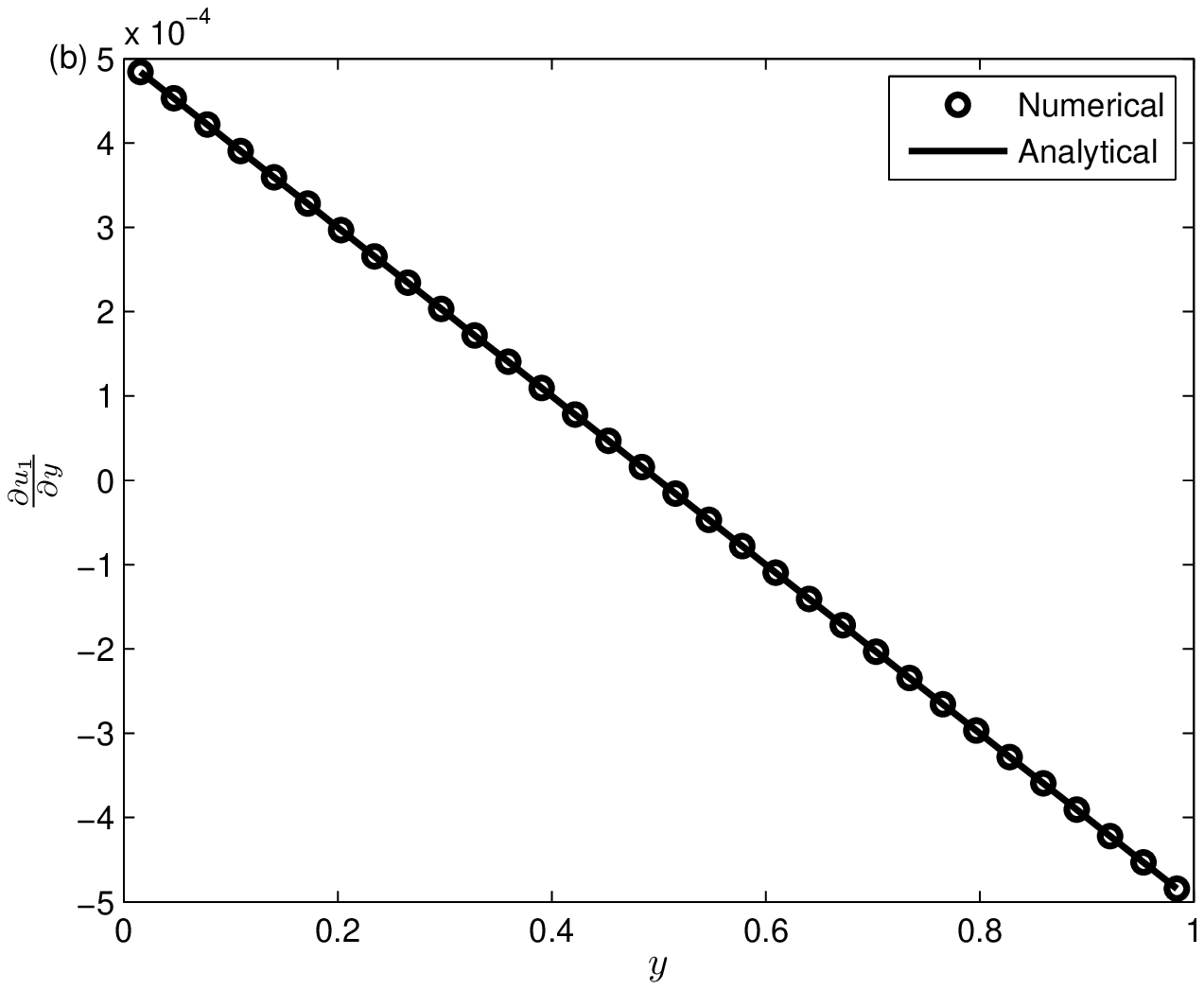}
\caption{The numerical and analytical solutions of the velocity gradient [(a): $\partial{u_{1}}/\partial x$, (b): $\partial{u_{1}}/\partial y$].}
\label{fig3}
\end{figure}

\begin{figure}
\centering
\includegraphics[width=0.6\textwidth]{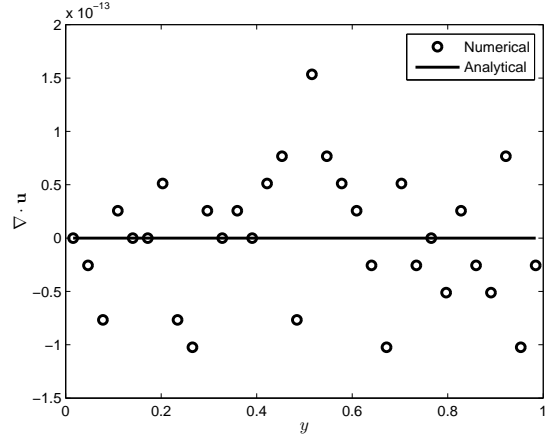}
\caption{The numerical and analytical solutions of the velocity divergence.}
\label{fig4}
\end{figure}

\begin{figure}
\centering
\includegraphics[width=0.6\textwidth]{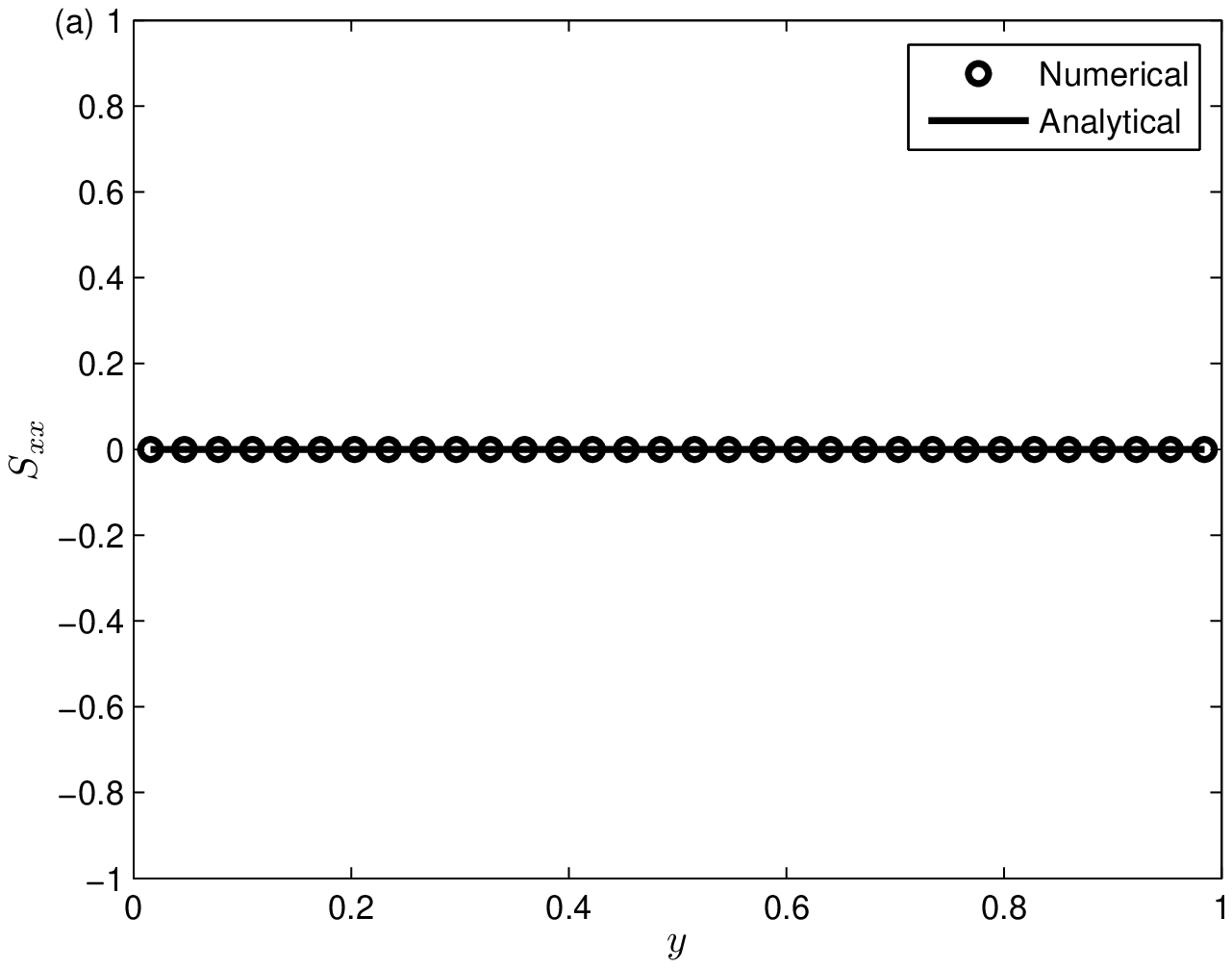}
\includegraphics[width=0.6\textwidth]{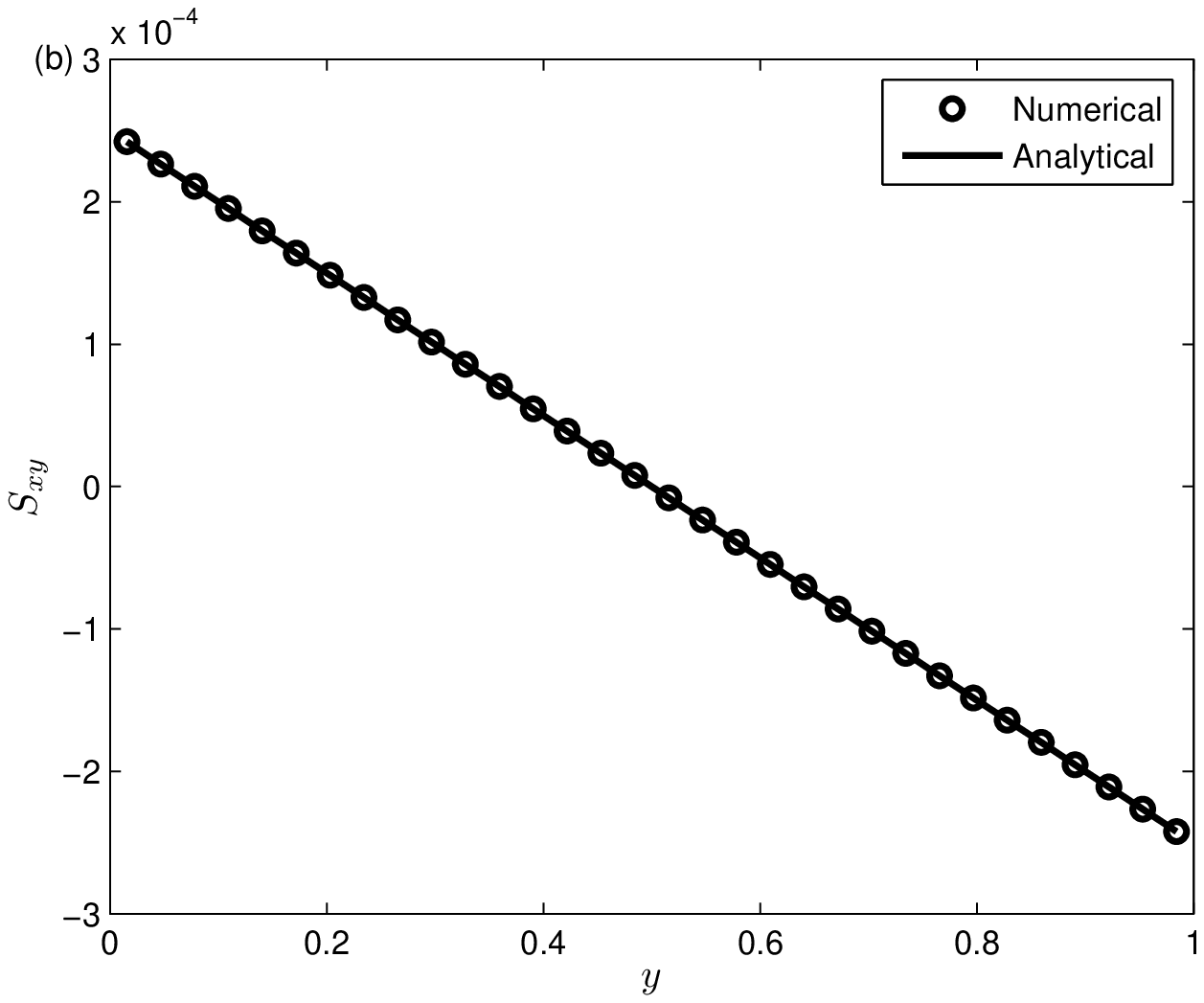}
\caption{The numerical and analytical solutions of the strain rate tensor [(a): $S_{xx}$, (b): $S_{xy}$].}
\label{fig5}
\end{figure}

\begin{figure}
\centering
\includegraphics[width=0.6\textwidth]{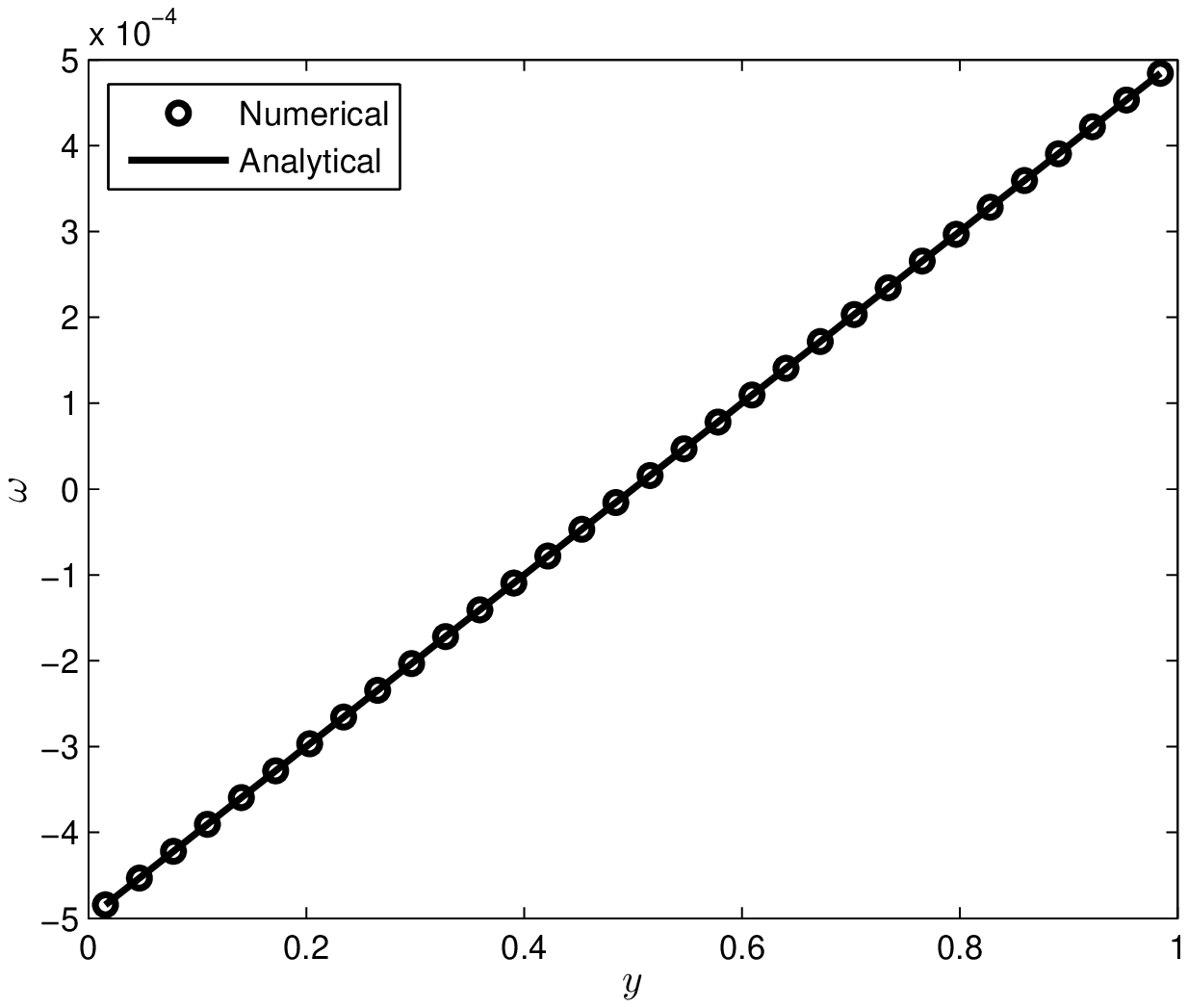}
\caption{The numerical and analytical solutions of the vorticity $\omega$.}
\label{fig6}
\end{figure}

We performed some simulations with the present MDF-LBM, and the lattice size $32\times32$ is adopted for the computational domain $[0, 1]\times [0, 1]$ with the periodic boundary condition in $x$ direction. We presented some numerical results in Figs. \ref{fig2}-\ref{fig6} where the kinematic viscosity is $\nu=0.001$, the relaxation parameters are set to be $s_{0}=1$, $s_{1}=1.2$ and $s_{2}=8(2-s_{1})/(8-s_{1})$ \cite{Cui2016, Wu2020}. As shown in these figures, the numerical results of the velocity $\mathbf{u}$, the components of velocity gradient ($\partial{u_{1}}/\partial x$ and $\partial{u_{1}}/\partial y$), velocity divergence $\nabla\cdot\mathbf{u}$, the components of strain rate tensor ($S_{xx}$ and $S_{xy}$), and the vorticity $\omega$ are in good agreement with analytical solutions. In addition, we also measured the relative errors of velocity $u_{1}$, the component of velocity gradient $\partial{u_{1}}/\partial y$, the component of strain rate tensor $S_{xy}$ and the vorticity $\omega$, and found that their values are about $E_{L^2} (u_{1})=5.9427\times10^{-4}$, $E_{L^2} (\partial{u_{1}}/\partial y)=9.7541\times10^{-17}$, $E_{L^2} (S_{xy})=9.7541\times10^{-17}$ and $E_{L^2} (\omega)=9.7541\times10^{-17}$. It should be noted that the component of velocity gradient $\partial{u_{1}}/\partial y$, the component of strain rate tensor $S_{xy}$ and the vorticity $\omega$ can achieve the machine precision. This is because the local schemes (\ref{eq:3-16}), (\ref{eq:3-17}) and (\ref{eq:3-18}) are of second-order accuracy, while the distributions of these physical variables are only linear. Finally, it is also observed from Fig. \ref{fig4} that the continuity equation $\nabla\cdot \mathbf{u}=0$ is preserved automatically.

\subsection{The simplified two-dimensional four-roll mill problem}
\begin{figure}
\centering
\includegraphics[width=0.6\textwidth]{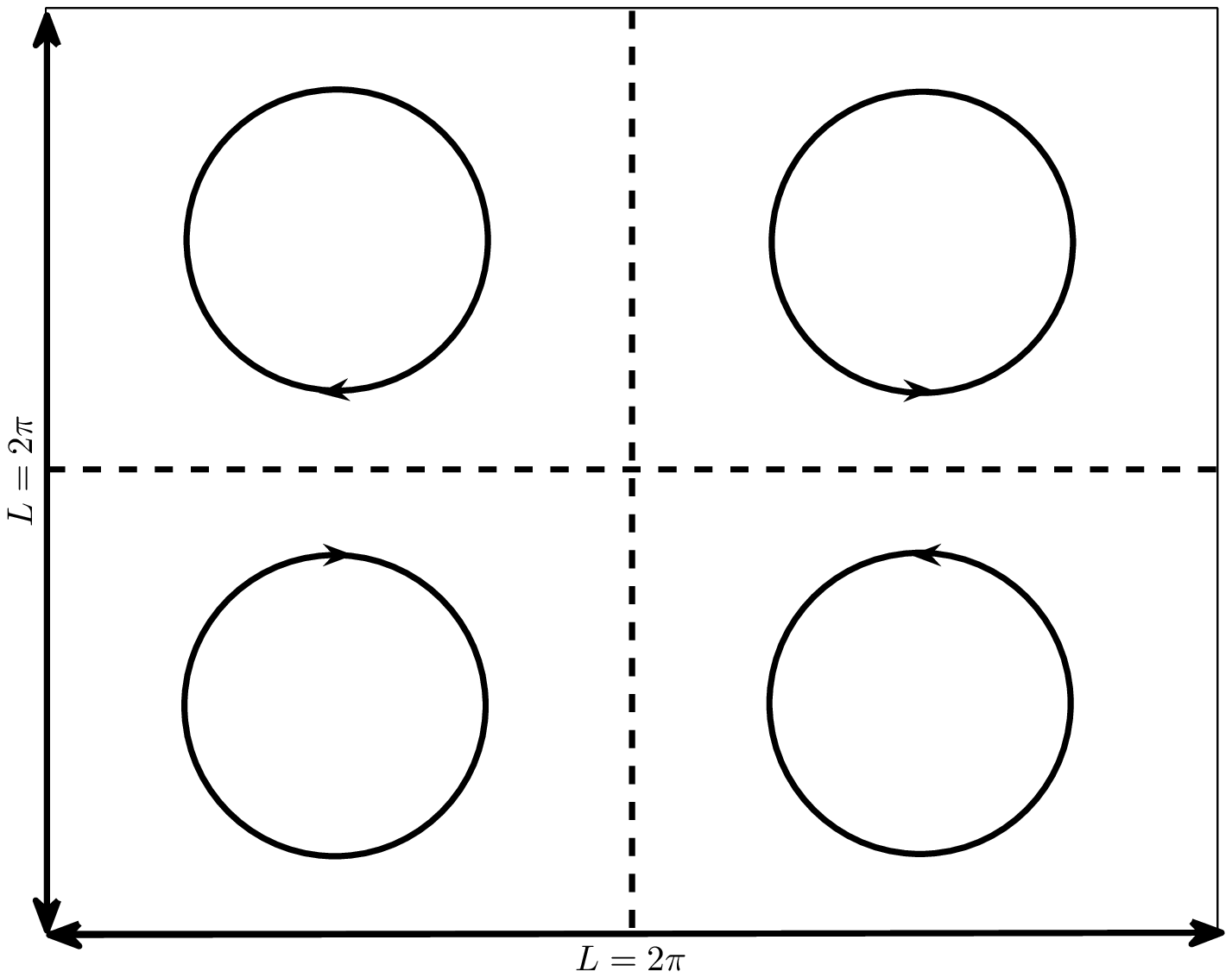}
\caption{The schematic of the two-dimensional four-roll mill problem.}
\label{fig7}
\end{figure}

The second example we used to test the present MDF-LBM is the simplified two-dimensional four-roll mil problem. The schematic of the problem is given in Fig. \ref{fig7} where the four rollers are replaced by a body force to drive fluid flow, the physical domain of the problem is $[0, 2\pi]\times [0, 2\pi]$ with the periodic boundary conditions in both $x$ and $y$ directions. As a benchmark problem, it has also been used to test the accuracy of the LBM \cite{Chai2012, Malaspinas2010} for the following two reasons. The first is that the problem has analytical solutions of velocity, velocity gradient, velocity divergence, strain rate tensor and vorticity. The second is that the boundary condition of the problem is periodic such that its effect on the numerical results can be excluded \cite{Chai2012}. For the incompressible flow driven by the following force,
\begin{subequations}\label{eq:4-2}
\begin{equation}
F_{1}=U_{0}^{2}\sin(x)\cos(x)+2\nu U_{0}\sin(x)\cos(y),
\end{equation}
\begin{equation}
F_{2}=U_{0}^{2}\sin(y)\cos(y)-2\nu U_{0}\sin(y)\cos(x),
\end{equation}
\end{subequations}
one can obtain the analytical solutions of velocity, velocity gradient, velocity divergence, stain rate tensor and vorticity,
\begin{subequations}\label{eq:4-2}
\begin{equation}
u_{1}=U_{0}\sin(x)\cos(y), \ u_{2}=-U_{0}\cos(x)\sin(y),
\end{equation}
\begin{equation}
\frac{\partial u_{1}}{\partial x}=U_{0}\cos(x)\cos(y)=-\frac{\partial u_{2}}{\partial y},\ \frac{\partial u_{1}}{\partial y}=-U_{0}\sin(x)\sin(y)=-\frac{\partial u_{2}}{\partial x},
\end{equation}
\begin{equation}
\nabla \cdot \mathbf{u}=\frac{\partial u_{1}}{\partial x}+\frac{\partial u_{2}}{\partial y}=0,
\end{equation}
\begin{equation}
S_{xx}=U_{0}\cos(x)\cos(y)=-S_{yy}, \ S_{xy}=S_{yx}=0,
\end{equation}
\begin{equation}
\omega=\frac{\partial u_{2}}{\partial x}-\frac{\partial u_{1}}{\partial y}=2U_{0}\sin(x)\sin(y).
\end{equation}
\end{subequations}

\begin{figure}
\centering
\includegraphics[width=0.6\textwidth]{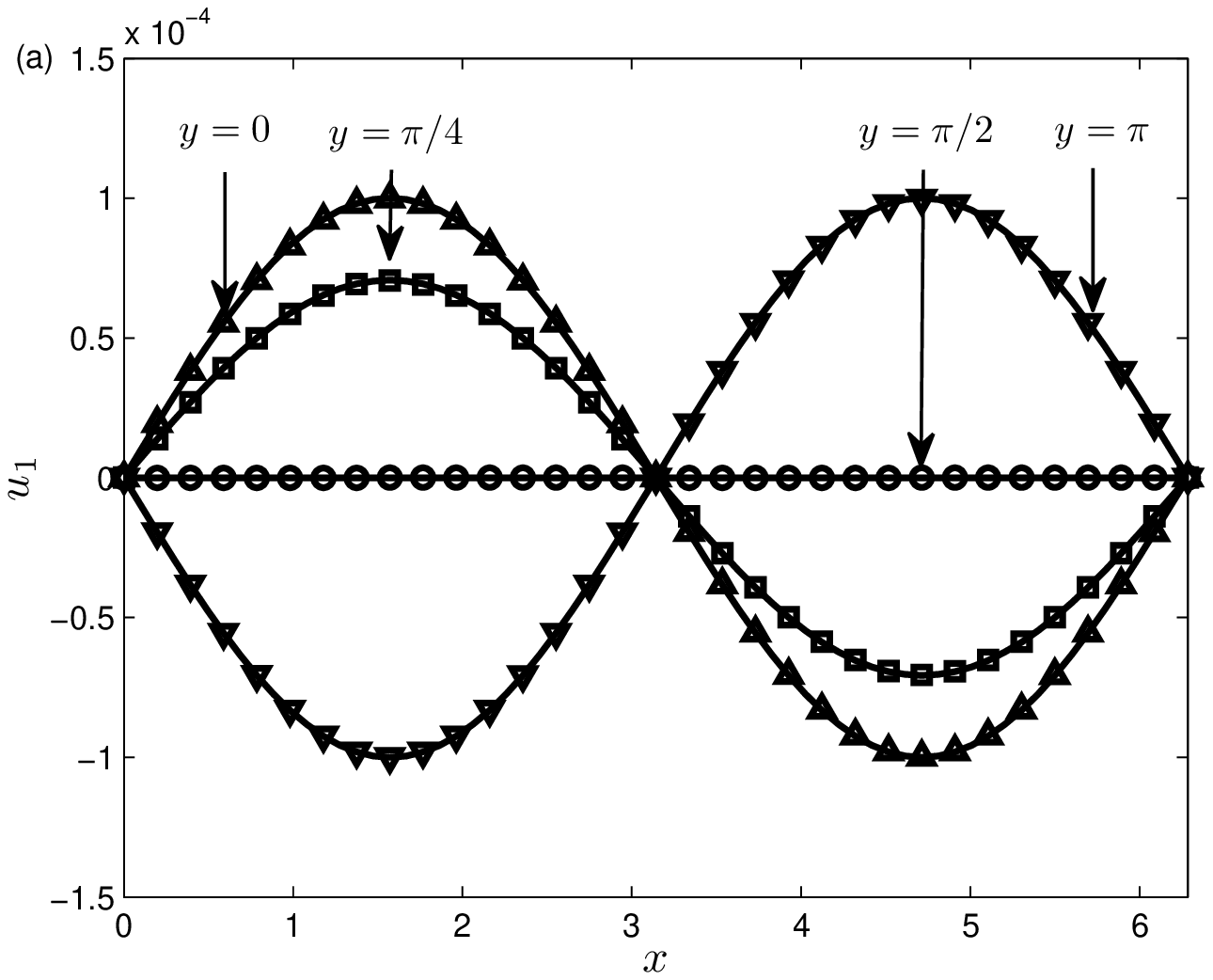}
\includegraphics[width=0.6\textwidth]{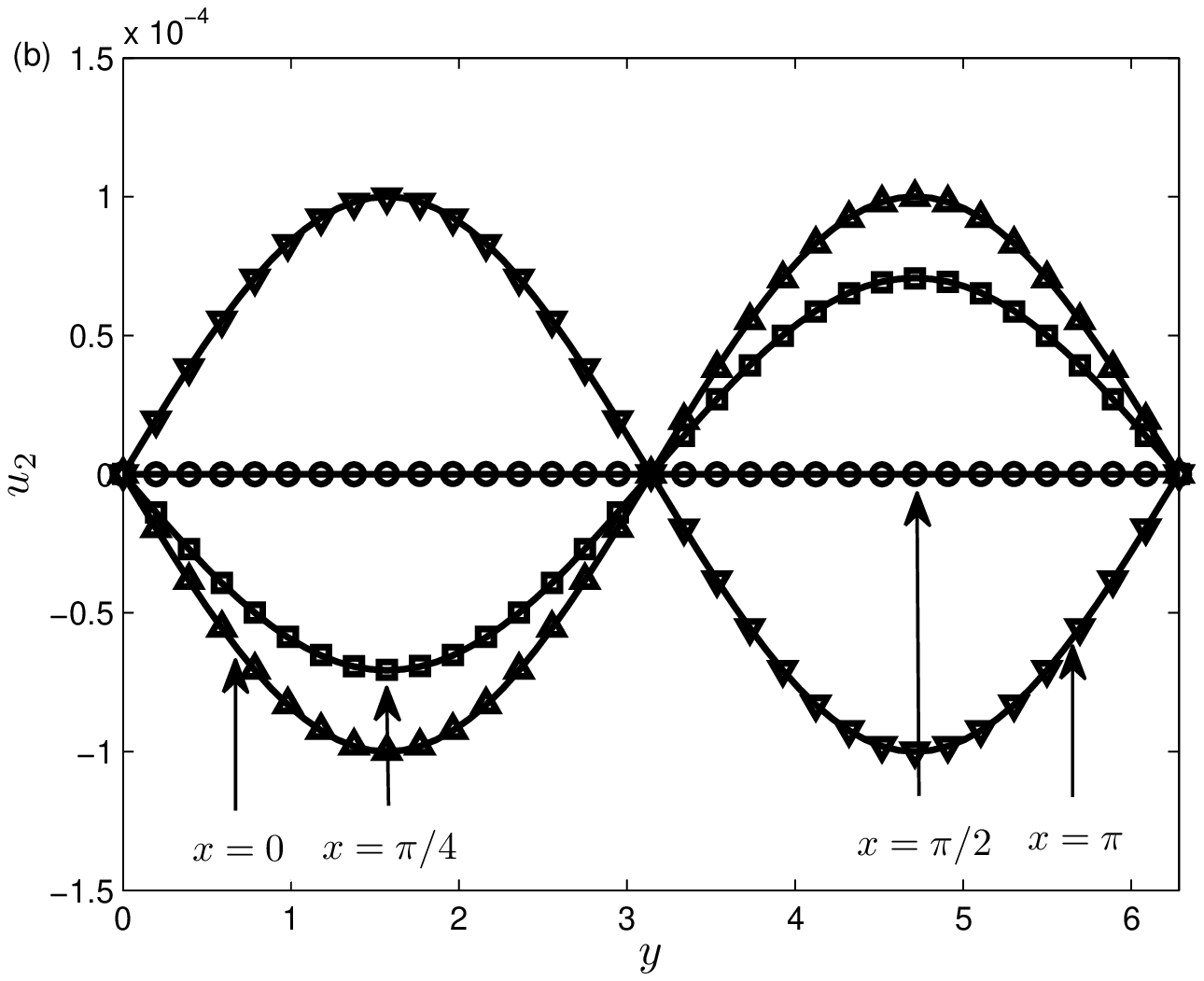}
\caption{The numerical and analytical solutions of velocity at different positions [(a): $u_{1}$, (b): $u_{2}$; solid line: analytical solution, symbol: numerical solution].}
\label{fig8}
\end{figure}

\begin{figure}
\centering
\includegraphics[width=0.6\textwidth]{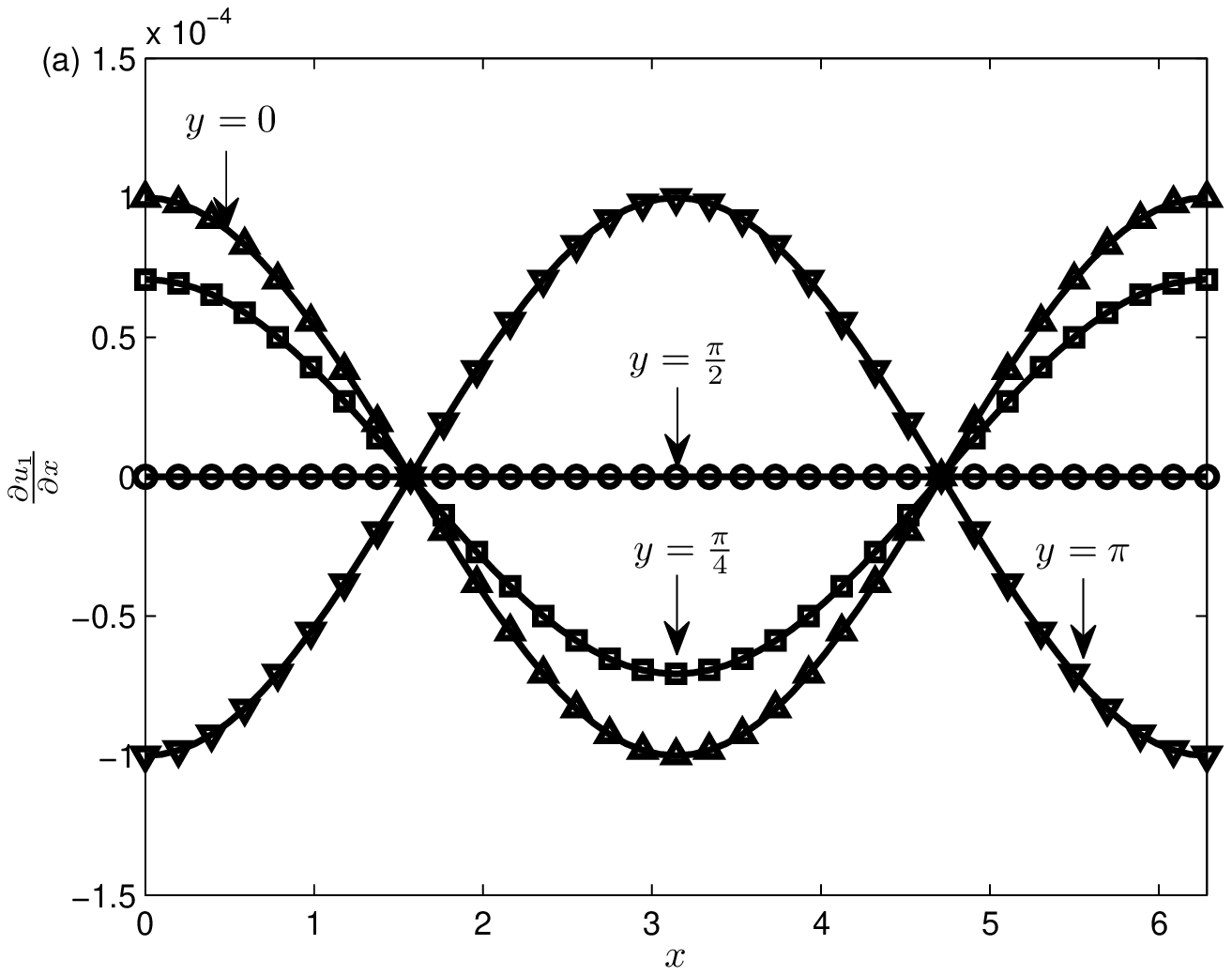}
\includegraphics[width=0.6\textwidth]{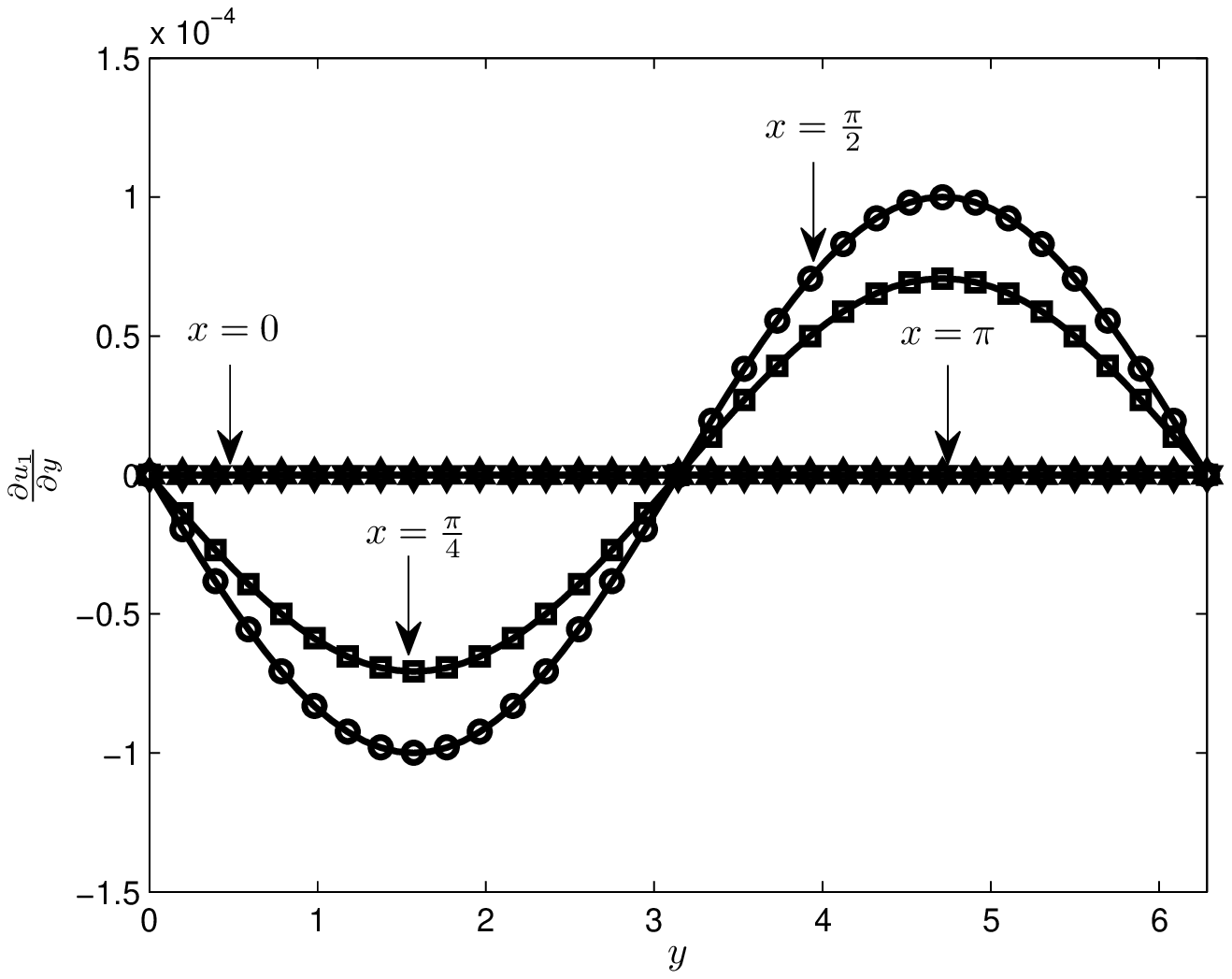}
\caption{The numerical and analytical solutions of velocity gradient at different positions [(a): $\partial u_{1}/\partial x$, (b): $\partial u_{1}/\partial y$; solid line: analytical solution, symbol: numerical solution].}
\label{fig9}
\end{figure}

\begin{figure}
\centering
\includegraphics[width=0.6\textwidth]{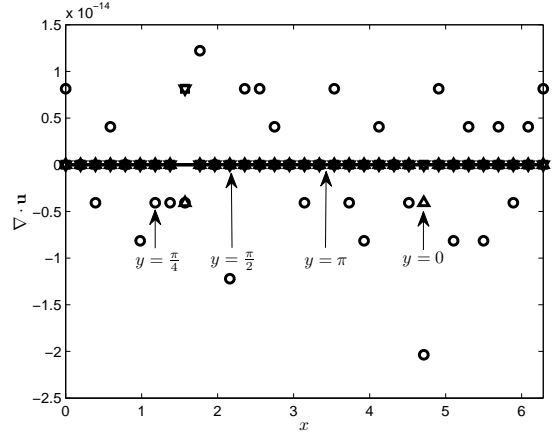}
\caption{The numerical and analytical solutions of the velocity divergence at different positions (solid line: analytical solution, symbol: numerical solution).}
\label{fig10}
\end{figure}

\begin{figure}
\centering
\includegraphics[width=0.6\textwidth]{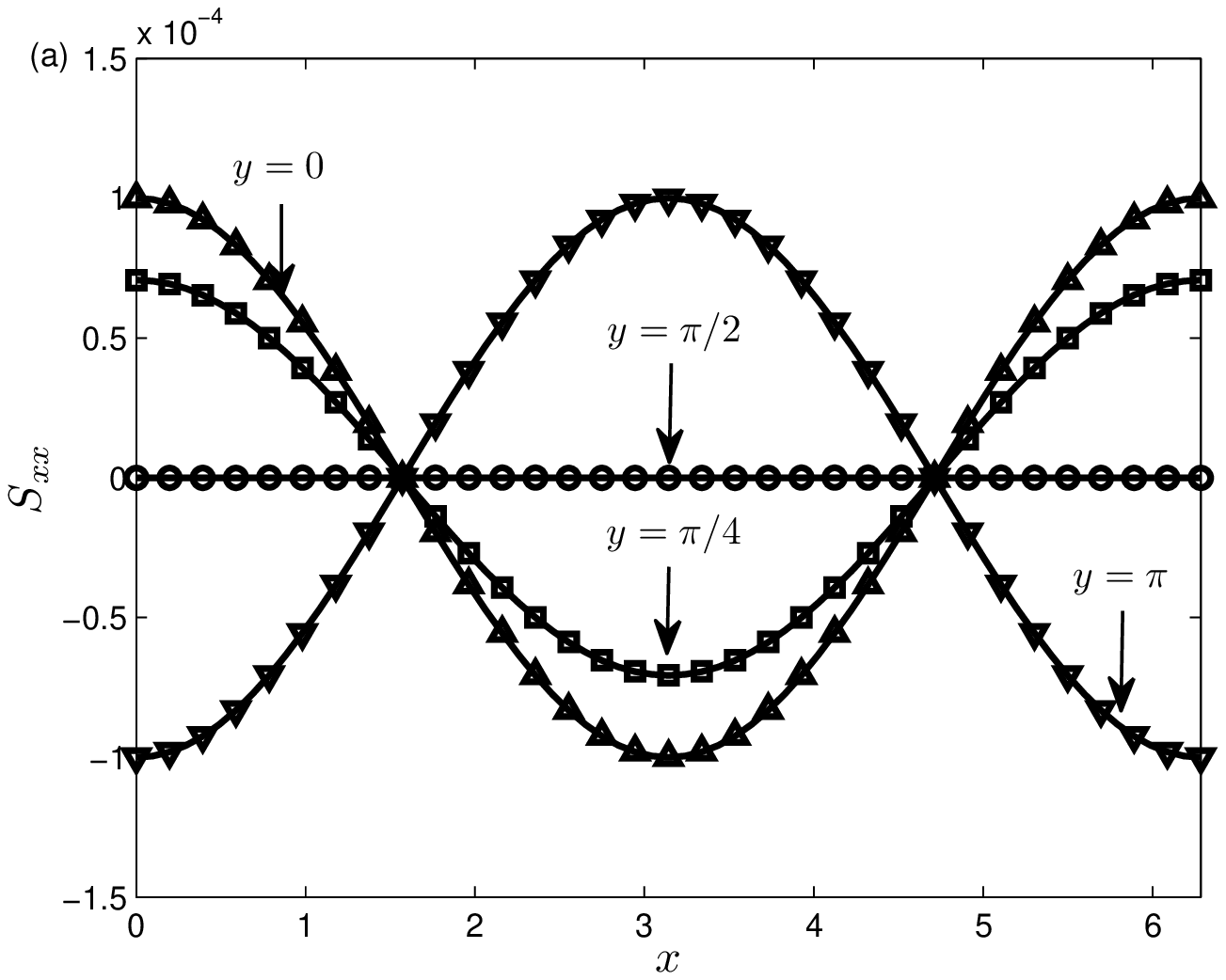}
\includegraphics[width=0.6\textwidth]{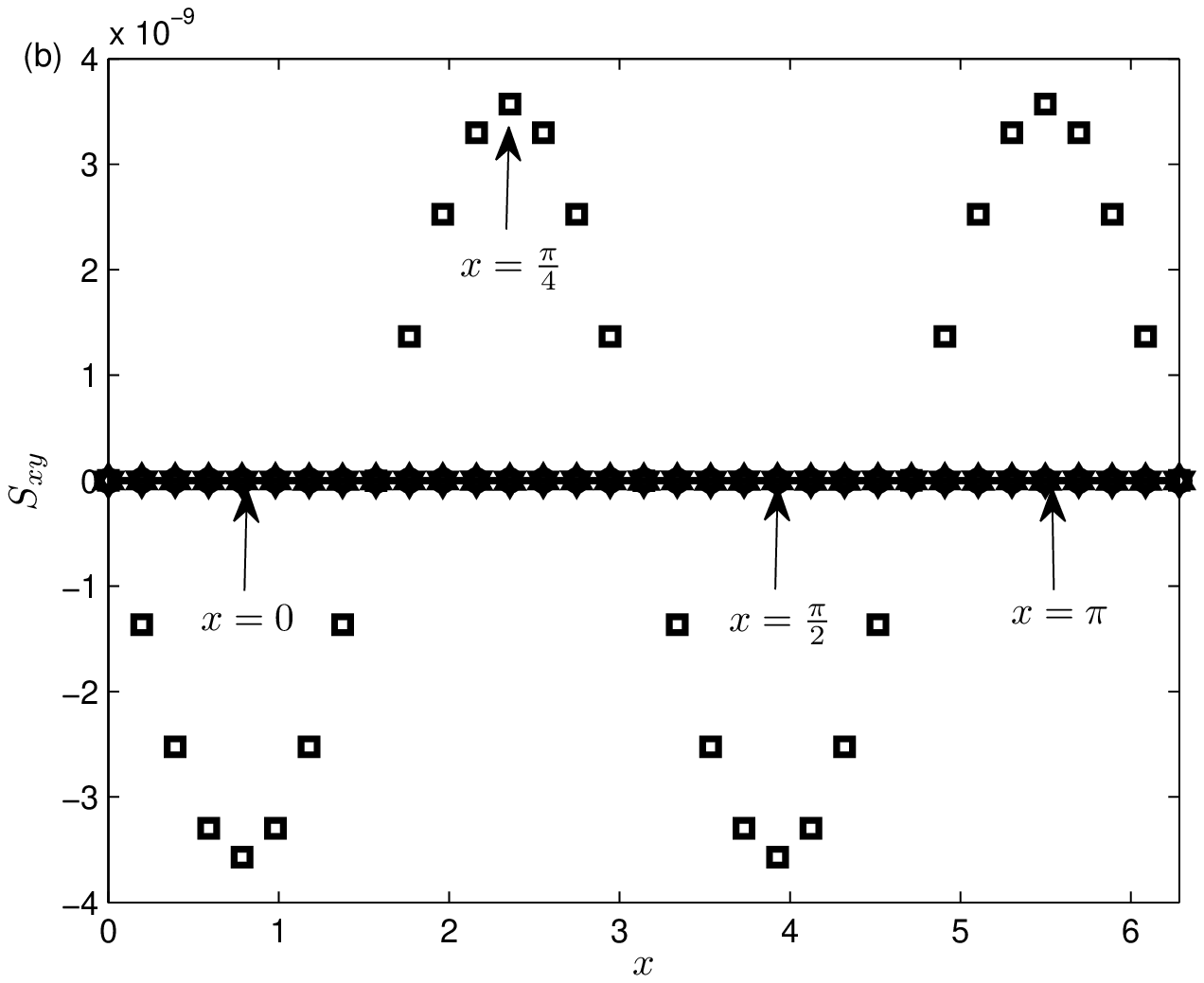}
\caption{The numerical and analytical solutions of the components of strain rate tensor at different positions [(a): $S_{xx}$, (b) $S_{yy}$; solid line: analytical solution, symbol: numerical solution].}
\label{fig11}
\end{figure}

\begin{figure}
\centering
\includegraphics[width=0.6\textwidth]{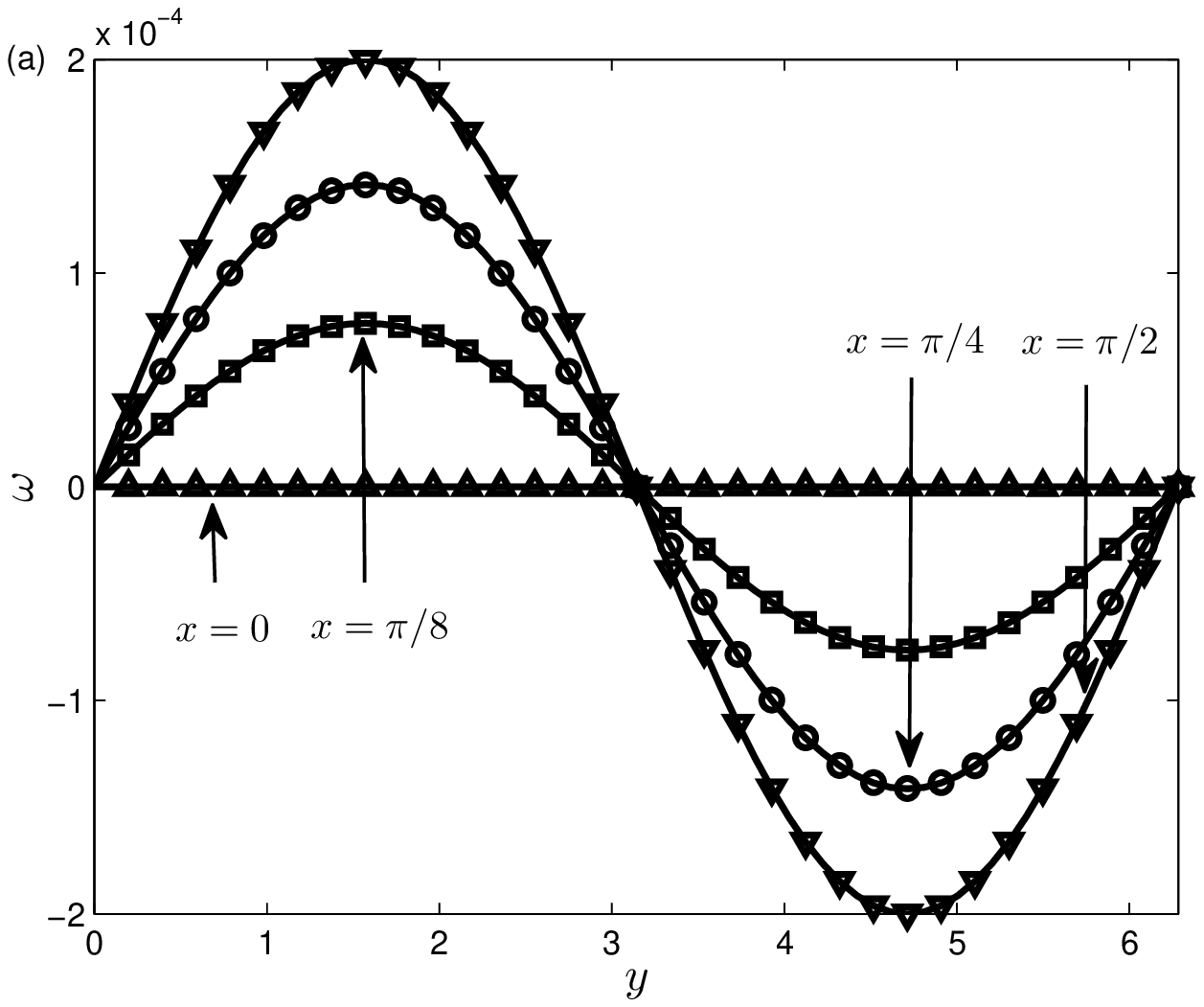}
\includegraphics[width=0.6\textwidth]{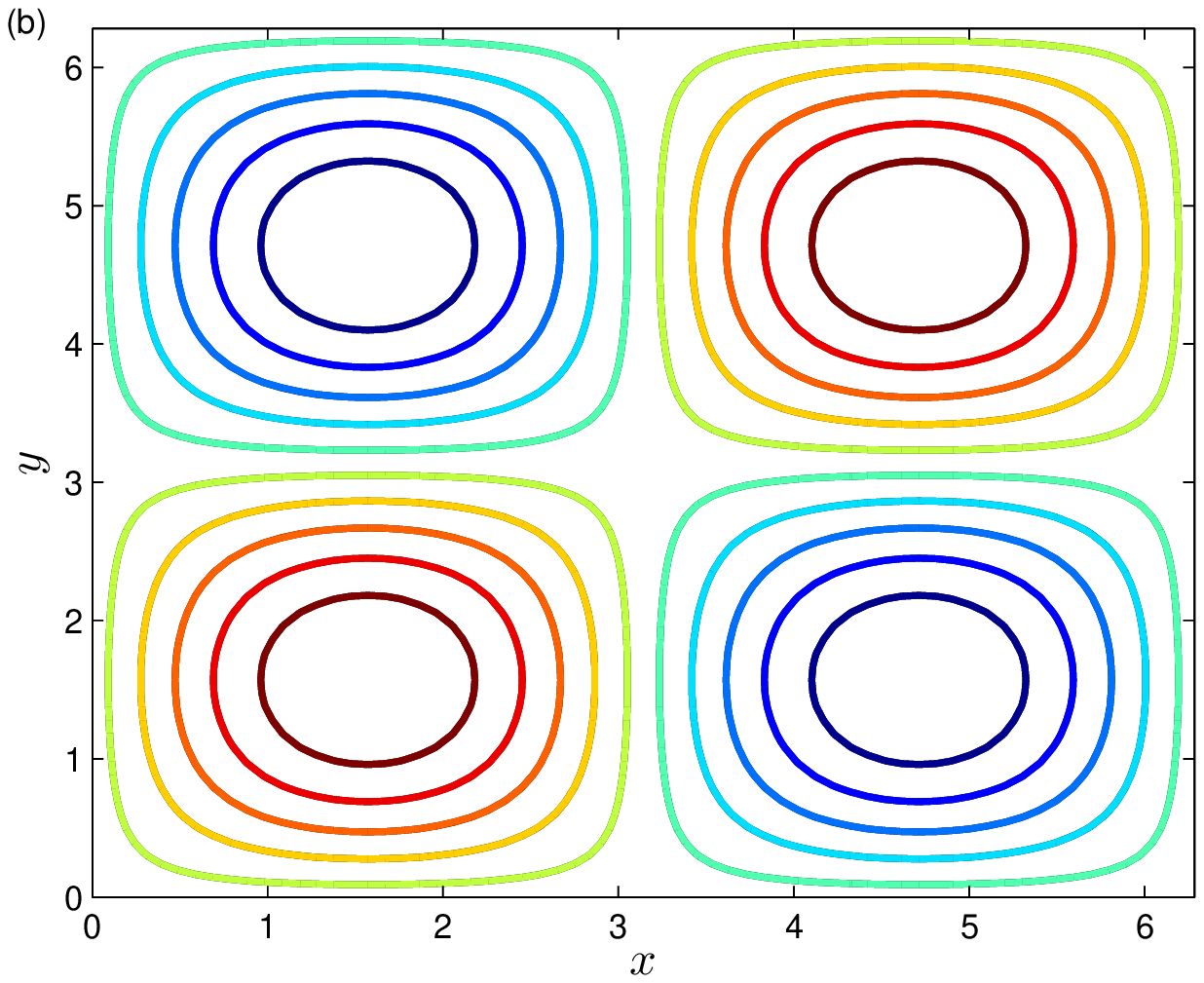}
\caption{The numerical and analytical solutions of vorticity at different positions (a) (solid line: analytical solution, symbol: numerical solution), and the contour lines of vorticity (b) (black line: analytical solution, colored line: numerical solution).}
\label{fig12}
\end{figure}

We carried out some numerical experiments with a lattice size $64\times 64$, and the results are shown in Figs. \ref{fig8}-\ref{fig12} where $U_{0}=0.0001$, $\nu=0.01$, the relaxation parameters are the same as those used in the first problem. As seen from these figures, the numerical results of velocity, velocity gradient, velocity divergence, strain rate tensor and vorticity are very close to the corresponding analytical solutions, and specially, from the contour lines of vorticity shown in Fig. \ref{fig12}(b), one can clearly observe that there are four vortices formed at the locations of the rollers.

\begin{figure}
\centering
\includegraphics[width=0.6\textwidth]{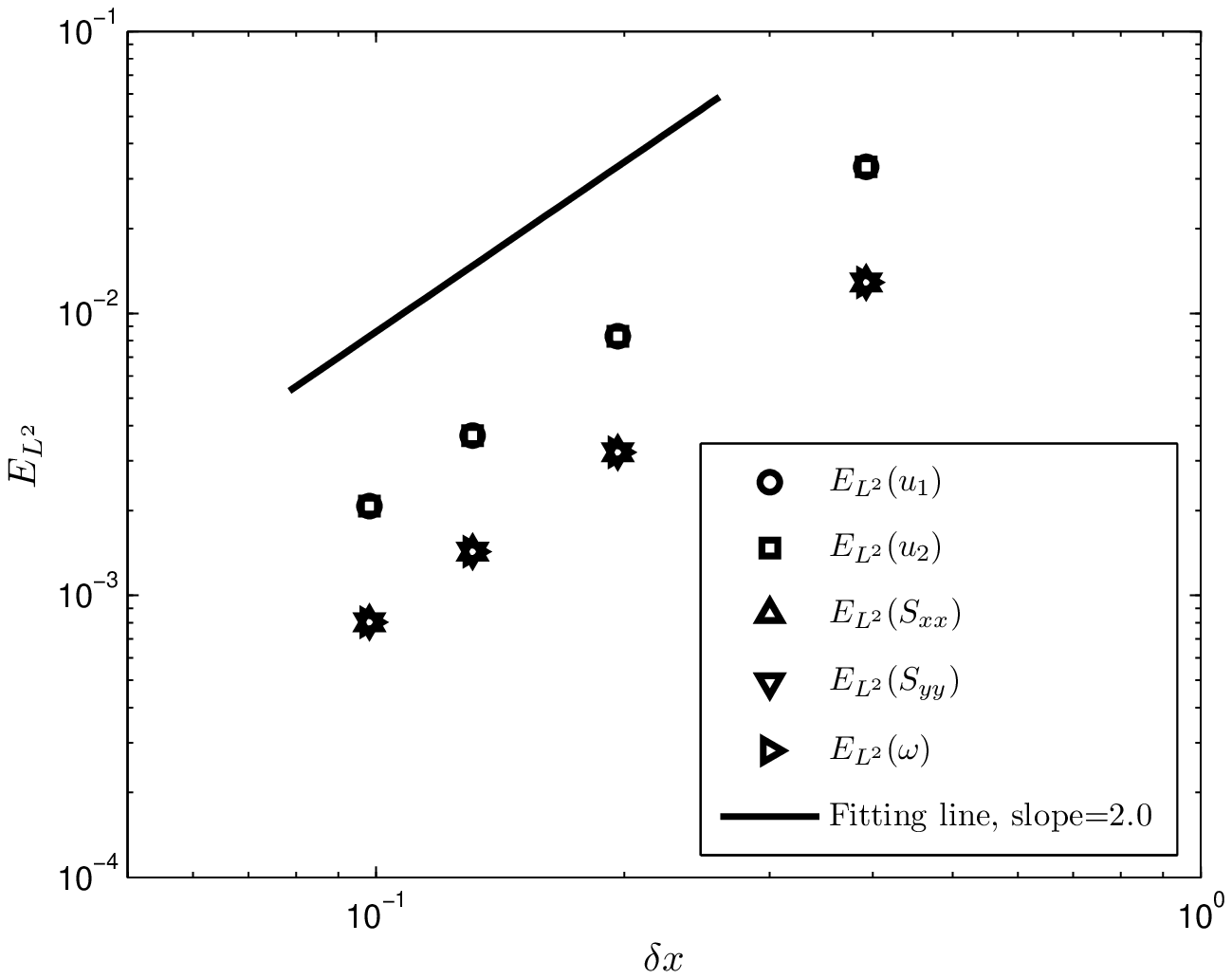}
\caption{The convergence rates of the MDF-LBM and the local schemes for the strain rate tensor and vorticity.}
\label{fig13}
\end{figure}

This problem is also used to test the convergence rates of the present MDF-LBM for velocity and the local schemes for the strain rate tensor and vorticity. To this end, we calculated the relative errors of velocity, the strain rate tensor and the vorticity at different lattice sizes ($N\times N=16 \times 16$, $32 \times 32$, $48 \times 48$ and $64 \times 64$), and plotted them in Fig. \ref{fig13}. From this figure, one can find that the present MDF-LBM and the local schemes are of second-order accuracy in space.

In addition, it is well known that the eigenvalue (or relaxation parameter) $s_{1}$ of the collision matrix $\mathbf{\Lambda}$ is a key parameter in the LBM, and may also affect the numerical results. Here we also conducted some simulations to test the effect of the relaxation parameter $s_{1}$, and presented the relative errors of the velocity, the strain rate tensor and the vorticity at three different values of the relaxation parameter $s_{1}$ in Table 1 where the viscosity and lattice size are fixed as $\nu=0.01$ and $64\times64$. As shown in this table, the relaxation parameter $s_{1}$ has no apparent influences on the numerical results, especially on the stain rate tensor and vorticity.

\begin{table*}
\caption{The relative errors of velocity, strain rate tensor and vorticity under different relaxation parameter $s_{1}$.}
\centering
\begin{tabular}{cccc}
\hline
  & $s_{1}=0.7$ & $s_{1}=1.2$ & $s_{1}=1.7$ \\ \hline
 $E_{L^2} (u_{1})$ & $3.9869\times10^{-3}$ & $2.0745\times10^{-3}$  & $1.2871\times10^{-3}$ \\
 $E_{L^2} (u_{2})$ & $3.9869\times10^{-3}$ & $2.0745\times10^{-3}$ & $1.2871\times10^{-3}$\\
 $E_{L^2} (S_{xx})$ & $8.0330\times10^{-4}$ & $8.0328\times10^{-4}$ & $8.0327\times10^{-4}$\\
  $E_{L^2} (S_{yy})$ & $8.0330\times10^{-4}$ & $8.0328\times10^{-4}$ & $8.0327\times10^{-4}$ \\
 $E_{L^2} (\omega)$ & $8.0333\times10^{-4}$ & $8.0330\times10^{-4}$ & $8.0330\times10^{-4}$ \\ \hline
\end{tabular}
\end{table*}

\subsection{The two-dimensional lid-driven cavity flow}
The last problem we considered is the two-dimensional lid-driven cavity flow, and the schematic of the problem is shown in Fig. 8 where the length of the square cavity is $L=1$. Compared to the previous problems, it is more complicated since there is no exact solution available. Although the geometry of the problem is very simple, the lid-driven cavity flow is of great scientific interest because it displays rich fluid mechanical phenomena, especially the complex vortex dynamics \cite{Shankar2000}. The flow in the square cavity is driven by the top moving wall with a constant velocity $(U_{1}, U_{2})^{\top}=(1, 0)^{\top}$, and a primary vortex in the center and some secondary vortices at the corners would be formed with the increase of the Reynolds number (Re$=LU_{1}/\nu$). Actually, the lid-driven cavity flow, as a classic benchmark problem, has also been widely used to test the capacity of the numerical methods \cite{Ghia1982, Hou1995, Botella1998, Erturk2006, Luo2011}.

In this part, we would conduct some numerical simulations of lid-driven cavity flows at different Reynolds numbers, and to ensure the incompressible condition to be valid, the discrete velocity $c=10$ is adopted to give a small Mach number (Ma$=U_{1}/c_{s}$). In our simulations, to obtain accurate results, the lattice size is set to be $512\times512$. We first presented the velocity profiles along the vertical and horizontal lines through geometric center of the cavity in Fig. 9 where Re$=100, 400$ and $1000$. As seen from this figure, the numerical results are in good agreement with the available data \cite{Ghia1982}. In addition, we also plotted the distributions of the pressure in Fig. 10, and found that these results qualitatively agree with the previous works \cite{Hou1995, Luo2011}. To show the complex dynamics of lid-driven cavity flows, we also plotted the streamlines and vorticity contours in Figs. 11 and 12. From these two figures, one can observe that a primary vortex in the center of the cavity and two secondary vortices at two bottom corners are formed. Simultaneously, it is also found that with the increase of the Reynolds number, the primary vortex would move towards the center of the cavity. To quantify these results, we measured the vorticities and locations of the primary and secondary vortices, and listed them in Table 2. As shown in this table, the present results are very close to those reported in some previous studies \cite{Ghia1982, Hou1995, Luo2011}.

\begin{figure}
\centering
\includegraphics[width=1.0\textwidth]{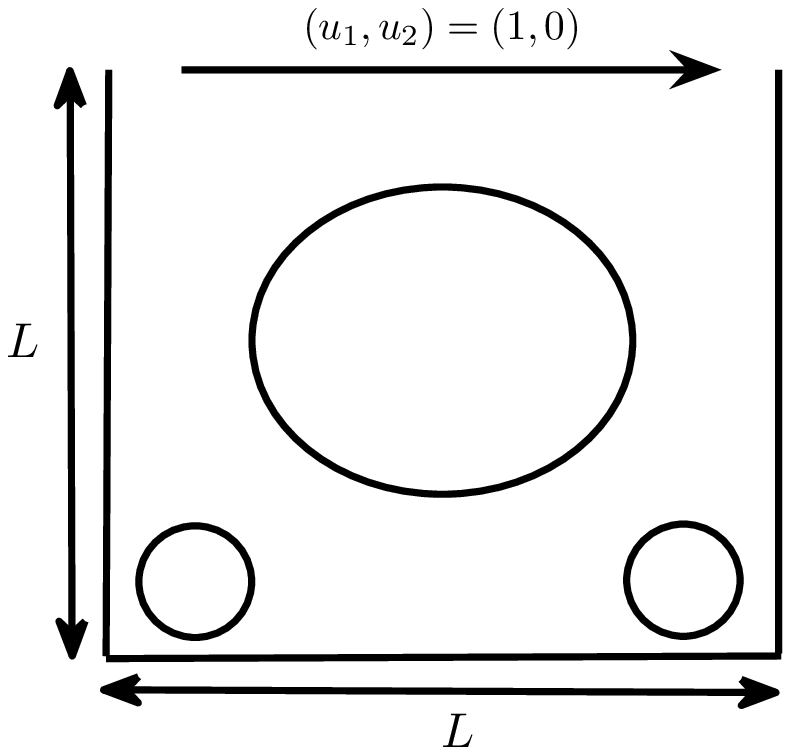}
\caption{The schematic of the two-dimensional lid-driven cavity flow.}
\label{fig14}
\end{figure}

\begin{figure}
\centering
\includegraphics[width=0.6\textwidth]{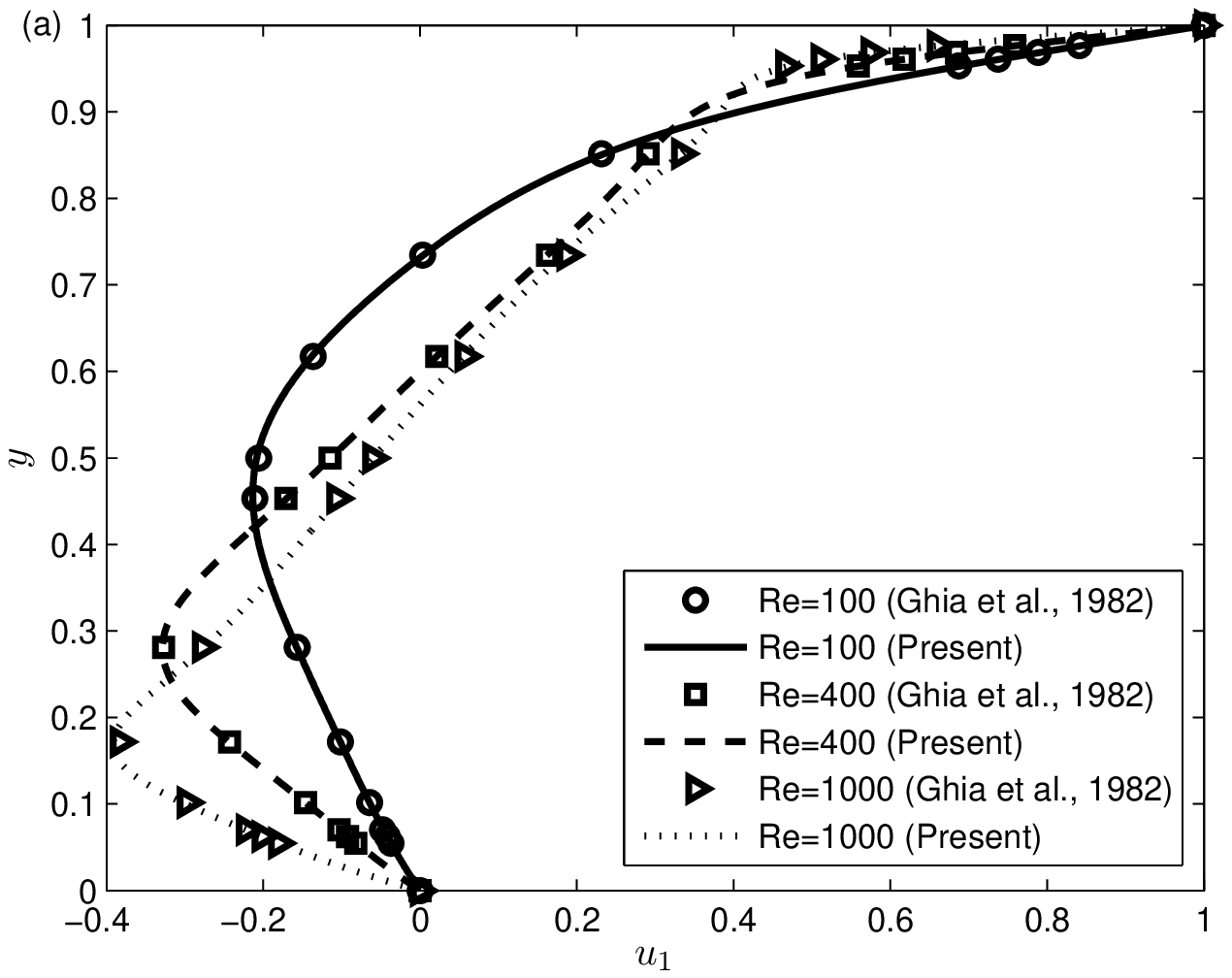}
\includegraphics[width=0.6\textwidth]{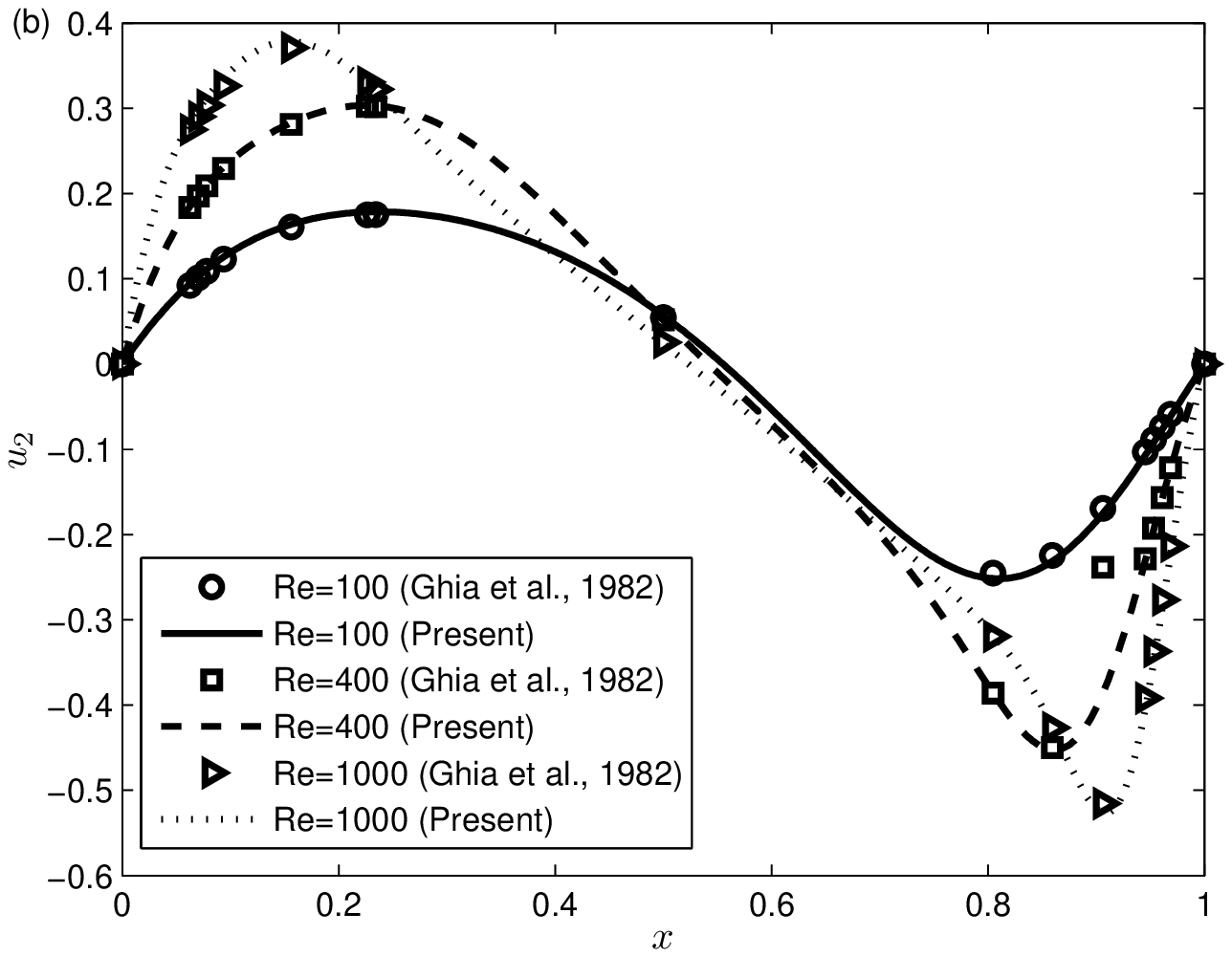}
\caption{The profiles of velocity along vertical and horizontal lines through geometric center of the cavity [(a): $u_{1}$ along vertical line; (b): $u_{2}$ along horizontal line].}
\label{fig15}
\end{figure}

\begin{figure}
\centering
\includegraphics[width=0.5\textwidth]{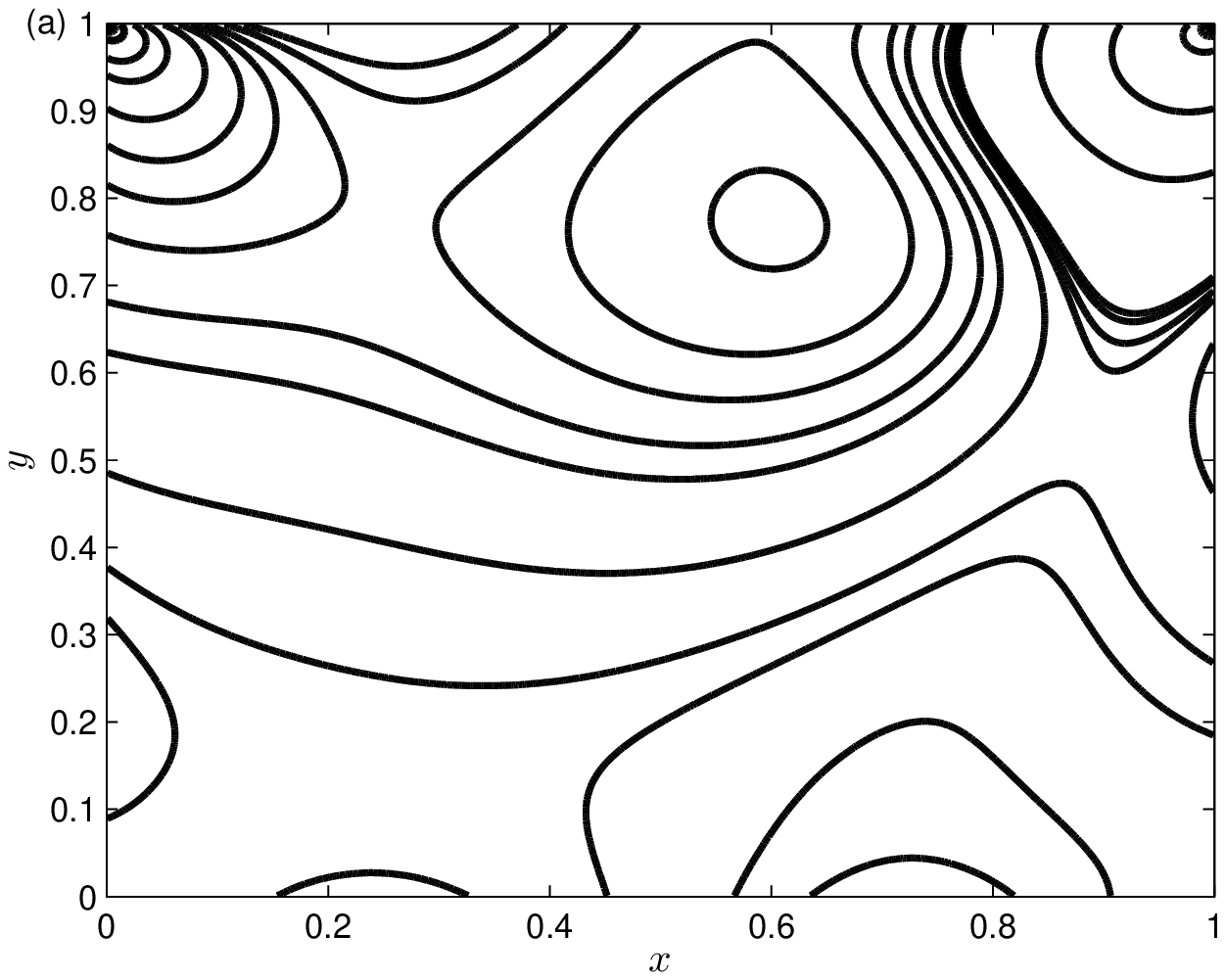}
\includegraphics[width=0.5\textwidth]{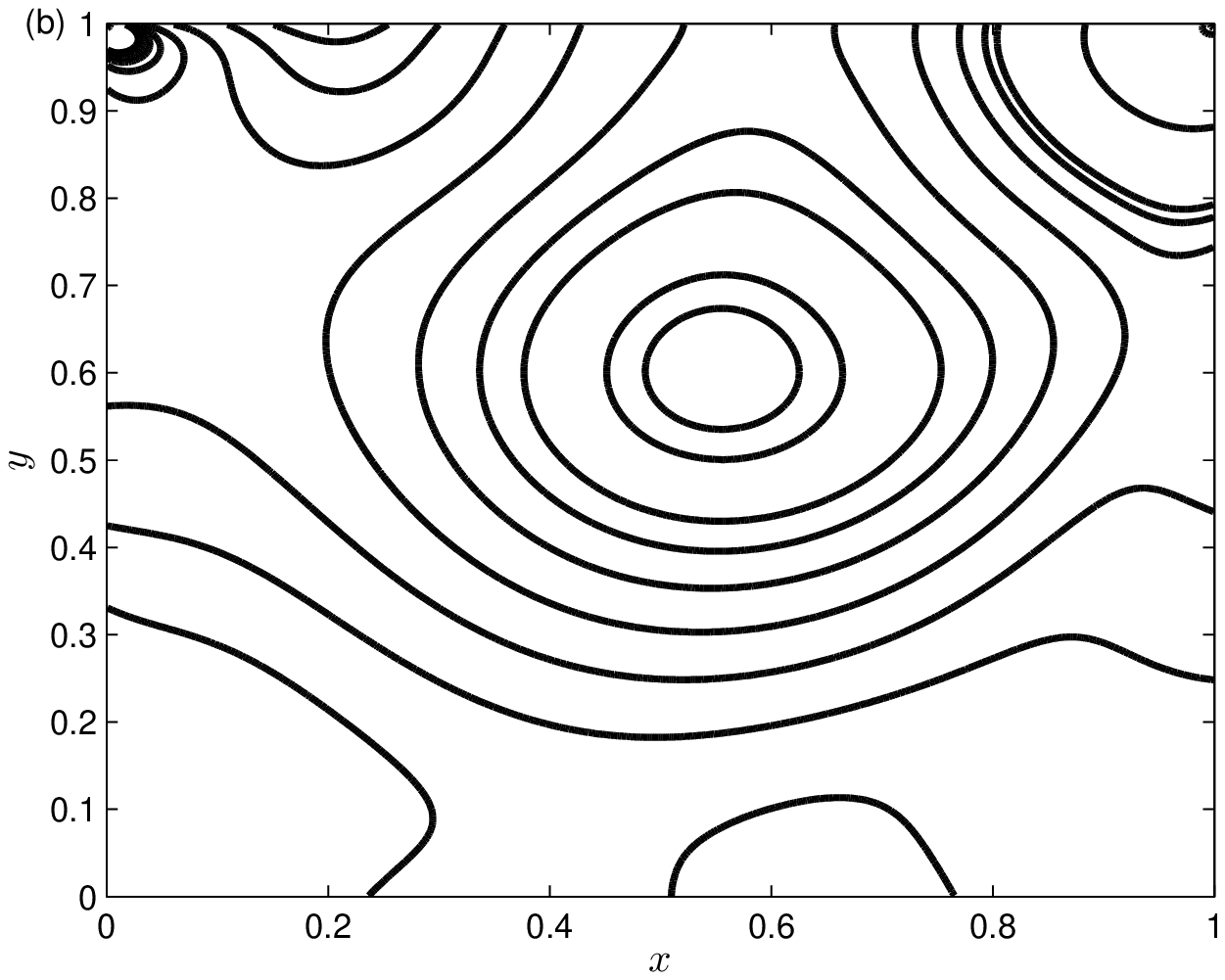}
\includegraphics[width=0.5\textwidth]{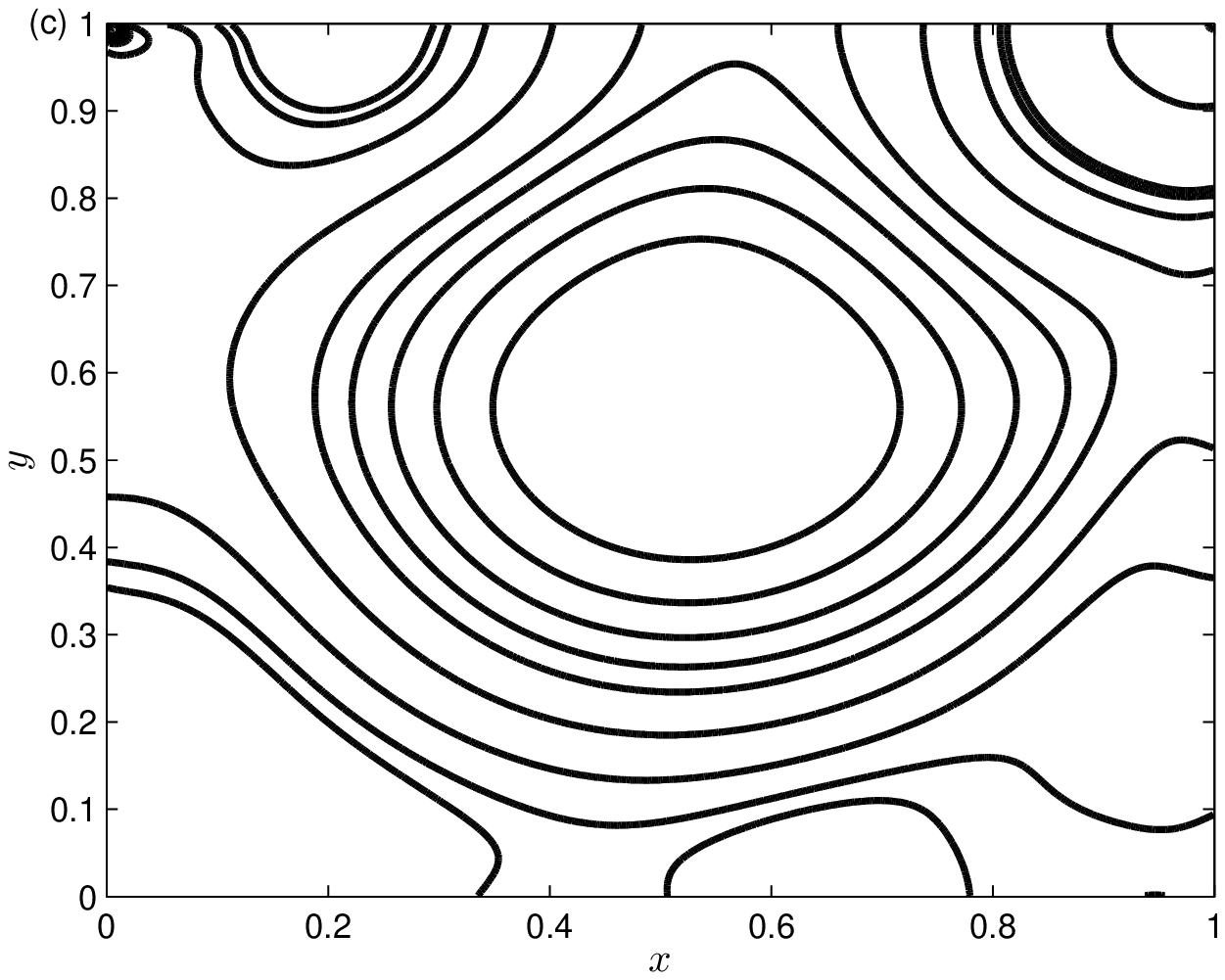}
\caption{The contours of pressure at Re=100 (a), 400 (b) and 1000 (c).}
\label{fig16}
\end{figure}

\begin{figure}
\centering
\includegraphics[width=0.6\textwidth]{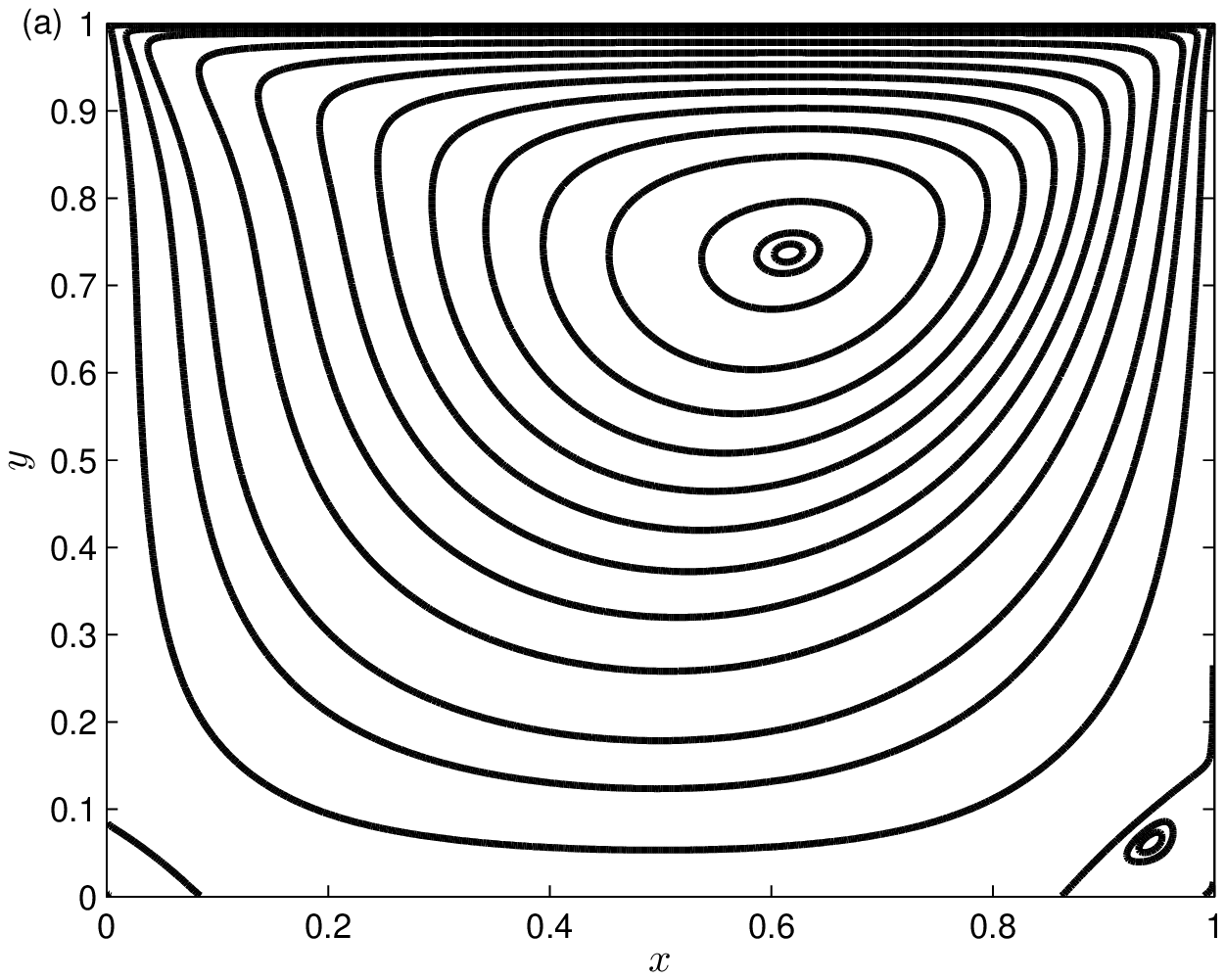}
\includegraphics[width=0.6\textwidth]{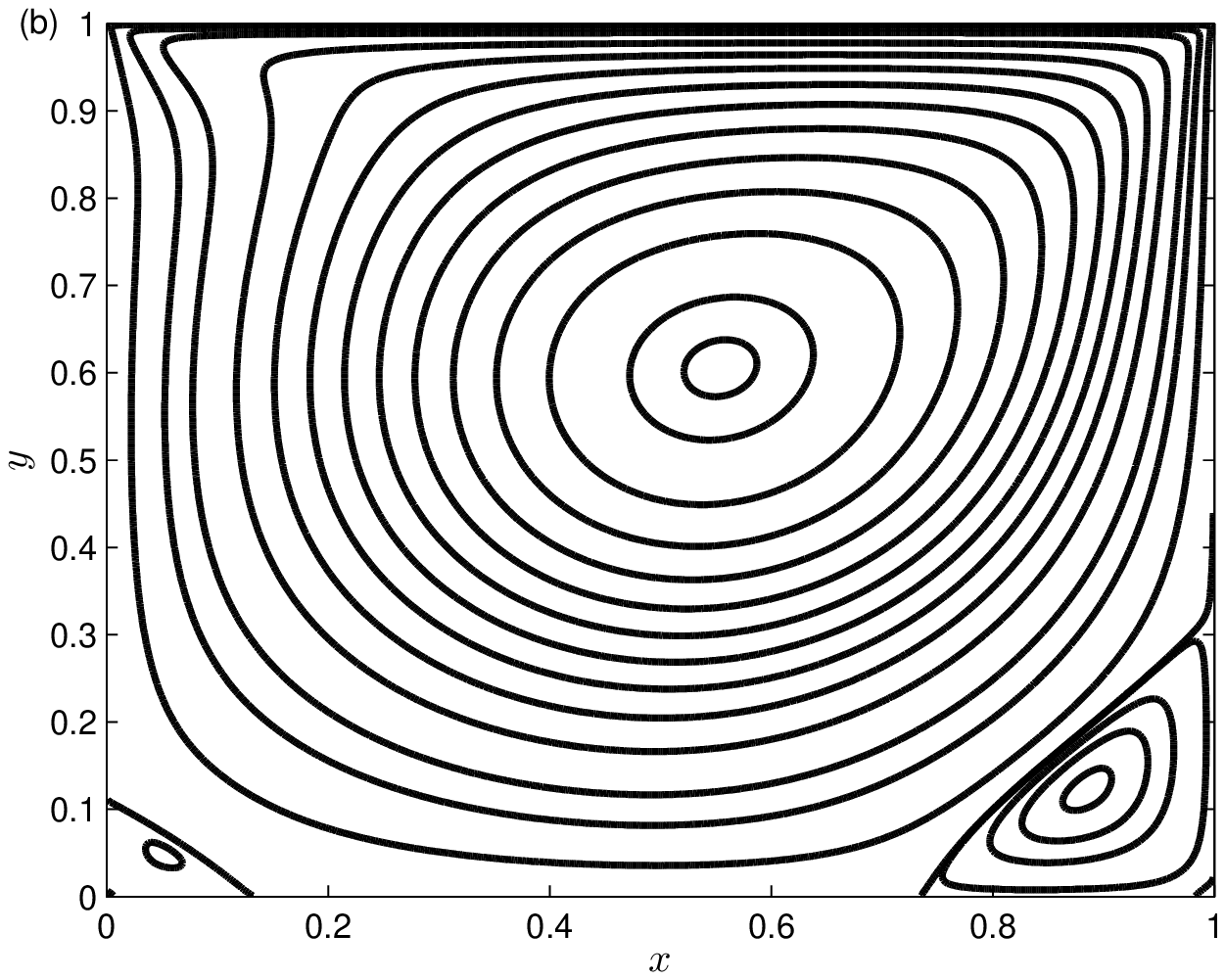}
\includegraphics[width=0.6\textwidth]{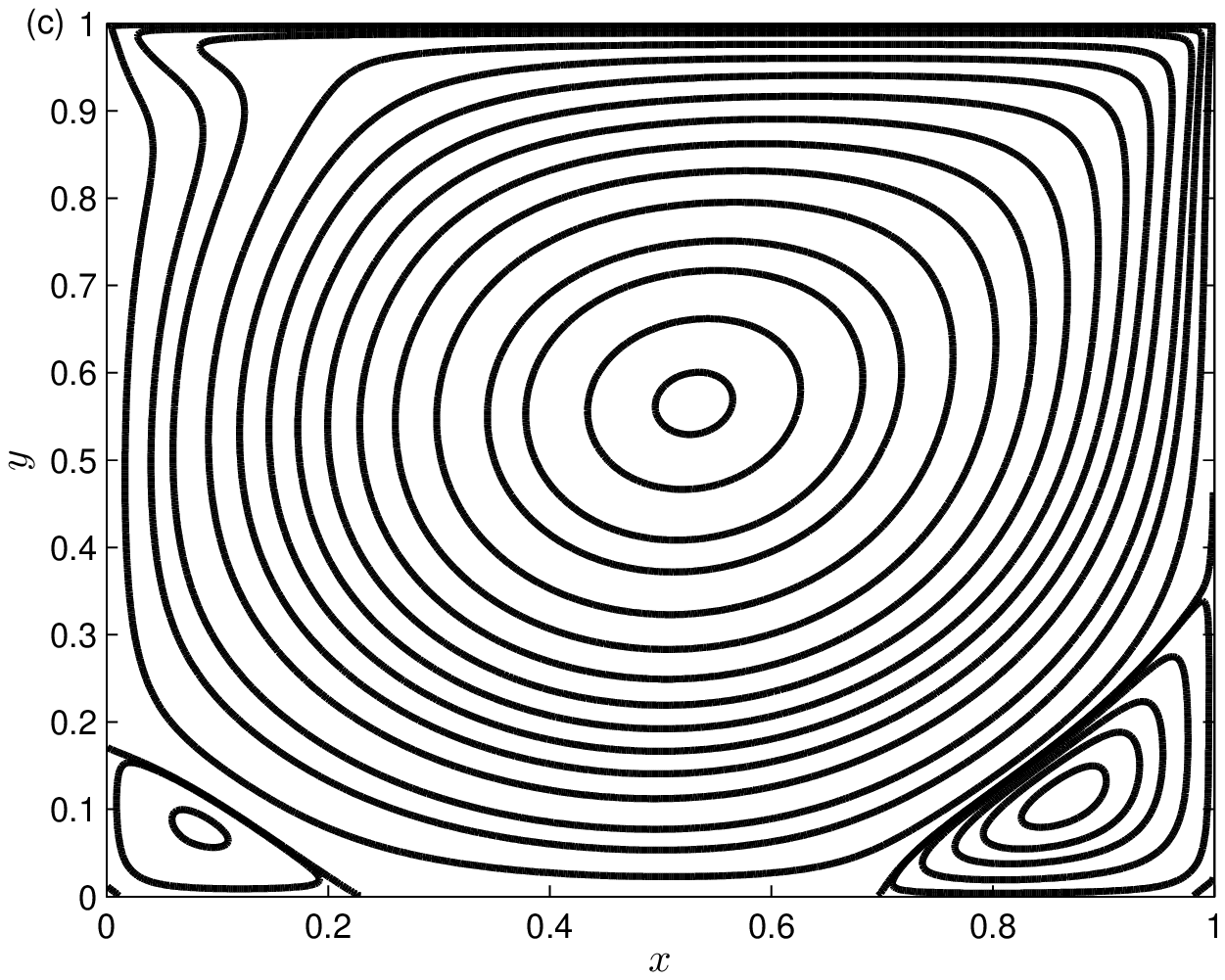}
\caption{The streamlines of lid-driven cavity flows at Re=100 (a), 400 (b) and 1000 (c).}
\label{fig17}
\end{figure}

\begin{figure}
\centering
\includegraphics[width=0.6\textwidth]{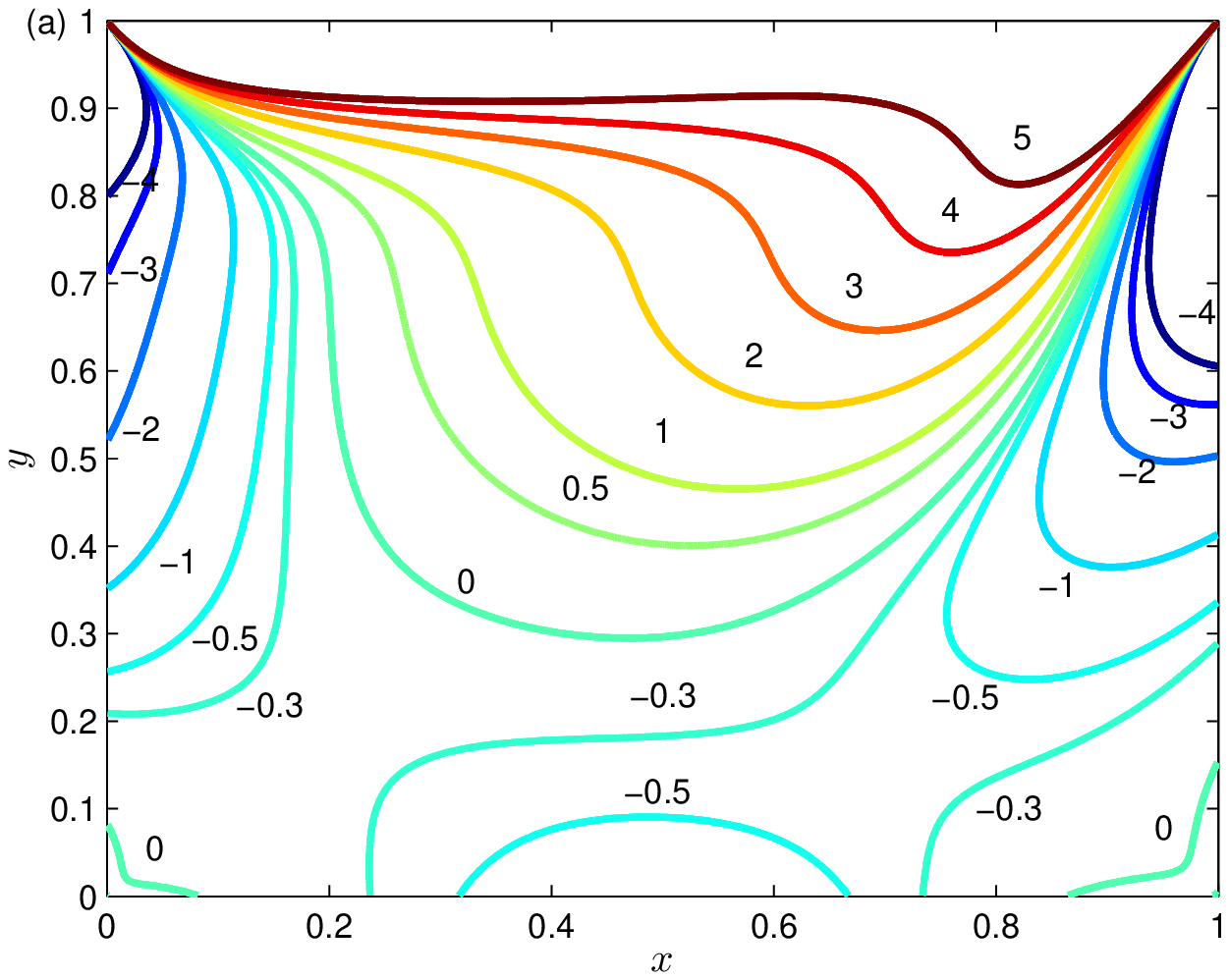}
\includegraphics[width=0.6\textwidth]{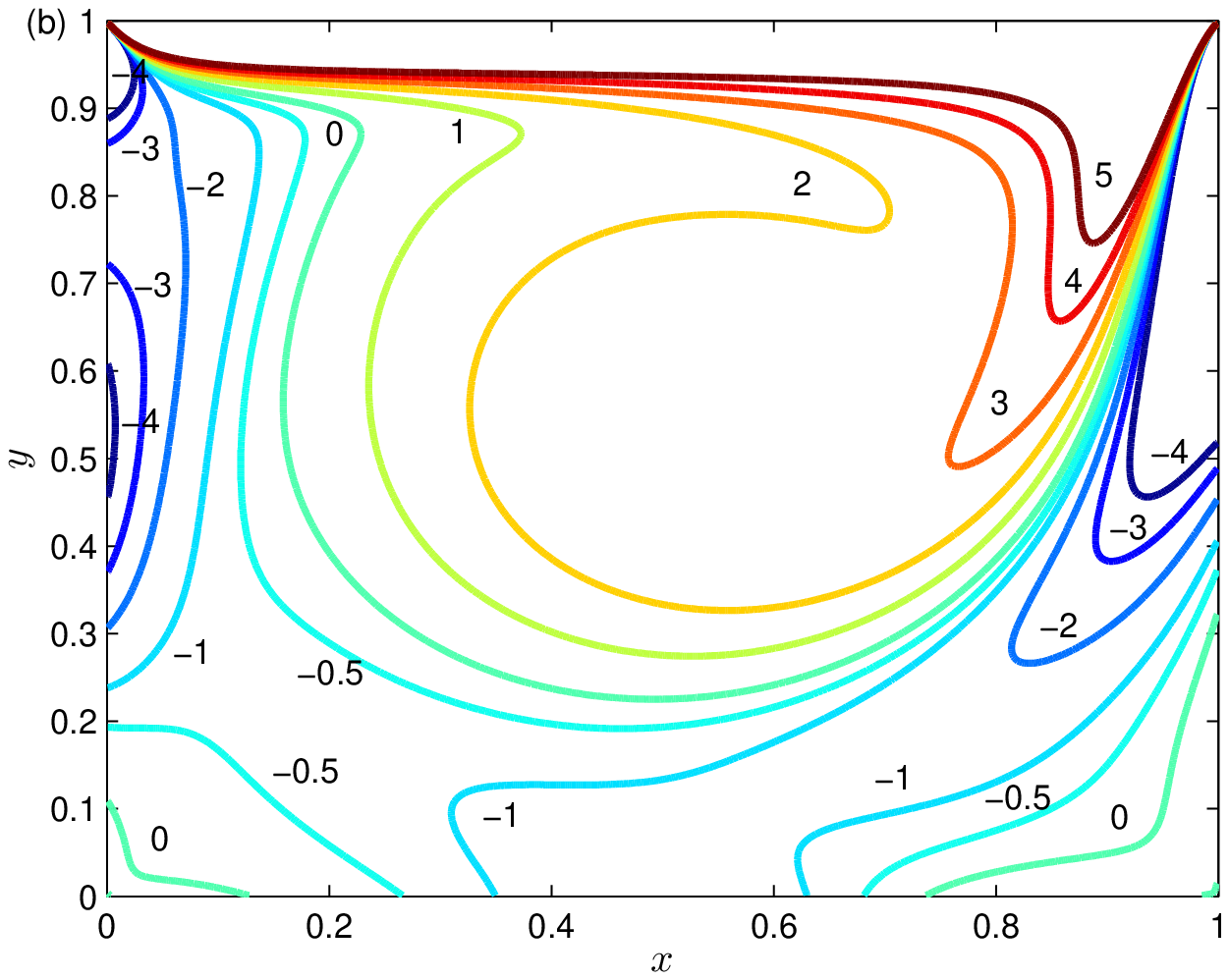}
\includegraphics[width=0.6\textwidth]{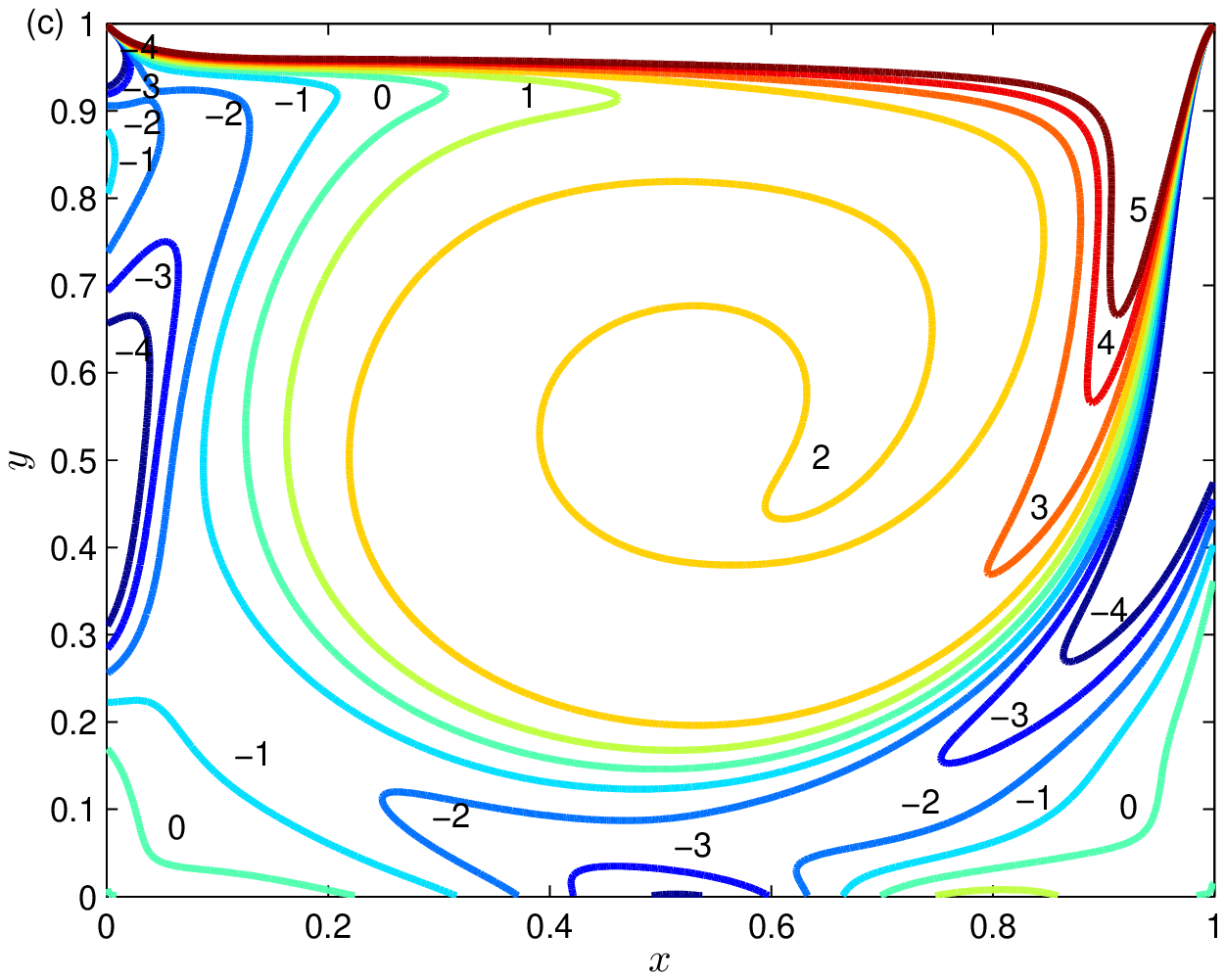}
\caption{The contours of the vorticity at Re=100 (a), 400 (b) and 1000 (c).}
\label{fig18}
\end{figure}

\begin{table*}
\caption{The vorticity ($\omega$) and location ($x$, $y$) of the vortex.}
\centering
\begin{tabular}{c|ccccc}
\hline
 Vortex & Reference & & Re=100 & Re=400 & Re=1000 \\ \hline
  \multirow{9}{*}{Primary vortex} & \multirow{3}{*}{Present} & $\omega$ & 3.1655  & 2.2797 & 1.9794\\
 & & $x$ & 0.6167  & 0.5540 & 0.5303\\
 & & $y$ & 0.7373  & 0.6053 & 0.5651\\ \cline{2-6}
 & \multirow{3}{*}{Ref. \cite{Ghia1982}} & $\omega$ & 3.1665  & 2.2947 & 2.0497\\
 & & $x$ & 0.6172  & 0.5547 & 0.5313\\
 & & $y$ & 0.7344  & 0.6055 & 0.5626\\ \cline{2-6}
 & \multirow{3}{*}{Ref. \cite{Hou1995}} & $\omega$ & 3.1348  & 2.2910 & 2.0760\\
 & & $x$ & 0.6196  & 0.5608 & 0.5333\\
 & & $y$ & 0.7373  & 0.6078 & 0.5647\\ \cline{2-6}
  & \multirow{3}{*}{Ref. \cite{Luo2011}} & $\omega$ & 3.1629  & 2.2950 & 2.0678\\
 & & $x$ & 0.6150  & 0.5546 & 0.5312\\
 & & $y$ & 0.7378  & 0.6053 & 0.5663\\ \hline
 \multirow{9}{*}{Bottom-left vortex} & \multirow{3}{*}{Present} & $\omega$ & -0.0149  & -0.0583 & -0.3514\\
 & & $x$ & 0.0343  & 0.0511 & 0.0833\\
 & & $y$ & 0.0345  & 0.0472 & 0.0778\\ \cline{2-6}
   & \multirow{3}{*}{Ref. \cite{Ghia1982}} & $\omega$ & -0.0155  & -0.0570 & -0.3618\\
 & & $x$ & 0.0313  & 0.0508 & 0.0859\\
 & & $y$ & 0.0391  & 0.0469 & 0.0781\\\cline{2-6}
 & \multirow{3}{*}{Ref. \cite{Hou1995}} & $\omega$ & -  & - & -\\
 & & $x$ & 0.0392  & 0.0549 & 0.0902\\
 & & $y$ & 0.0353  & 0.0510 & 0.0784\\ \cline{2-6}
 & \multirow{3}{*}{Ref. \cite{Luo2011}} & $\omega$ & 0.0145  & 0.0602 & 0.3557\\
 & & $x$ & 0.0341  & 0.0517 & 0.0828\\
 & & $y$ & 0.0341  & 0.0478 & 0.0789\\ \hline
 \multirow{9}{*}{Bottom-right vortex} & \multirow{3}{*}{Present} & $\omega$ & -0.0352  & -0.4486 & -1.0690\\
 & & $x$ & 0.9425  & 0.8855 & 0.8650\\
 & & $y$ & 0.0619  & 0.1222 & 0.1120\\ \cline{2-6}
   & \multirow{3}{*}{Ref. \cite{Ghia1982}} & $\omega$ & -0.0331  & -0.4335 & -1.1547\\
 & & $x$ & 0.9453  & 0.8906 & 0.8594\\
 & & $y$ & 0.0625  & 0.1250 & 0.1094\\ \cline{2-6}
 & \multirow{3}{*}{Ref. \cite{Hou1995}} & $\omega$ & - & - & -\\
 & & $x$ & 0.9451  & 0.8902 & 0.8667\\
 & & $y$ & 0.0627  & 0.1255 & 0.1137\\ \cline{2-6}
  & \multirow{3}{*}{Ref. \cite{Luo2011}} & $\omega$ & -0.0348 & -0.4451 & -1.1039\\
 & & $x$ & 0.9425  & 0.8560 & 0.8645\\
 & & $y$ & 0.0614  & 0.1218 & 0.1121\\  \hline
\end{tabular}
\end{table*}
\section{Conclusions}

In this work, a MDF-LBM coupled with the MRT model is developed for incompressible NSEs. To do this, the NSEs are first reformulated into a convection-diffusion system, and then the MDF-LBM is proposed for the convection-diffusion system. What is more, besides the macroscopic pressure and velocity, we also proposed some local schemes for the velocity gradient, velocity divergence, strain rate tensor, shear stress and vorticity in the framework of MDF-LBM. To test the capacity of the MDF-LBM and local schemes, three benchmark problems, including the two-dimensional Poiseuille flow, the simplified four-roll mill problem and the lid-driven cavity flow, are considered. The numerical results show that the present MDF-LBM and local schemes are efficient, and also have a second-order convergence rate in space.

It should be noted that compared to the classic LBM for two-dimensional incompressible NSEs where the D2Q9 lattice model should be used, the MDF-LBM is more flexible since the D2Q4, D2Q5 or D2Q9 lattice model can be adopted to give correct incompressible NSEs. Additionally, in the MDF-LBM for convection-diffusion-system based NSEs, some physical variables (e.g., the velocity gradient, velocity divergence, the strain rate tensor, the shear stress and the vorticity) can be computed locally through the first-order moments of the non-equilibrium distribution function, while in the commonly used single LBM for incompressible NSEs, usually only the velocity divergence, strain rate tensor and the shear stress can be determined locally by the the second-order moments of the non-equilibrium distribution function.

Finally, we would also like to point out that the present MDF-LBM can be also extended to study the thermal flows and multiphase fluid system governed by the incompressible NSEs and CDE, which would be considered in another work.

\section*{Acknowledgements}
This work was financially supported by the National Natural Science Foundation of China (Grants No. 12072127 and No. 51836003).

\end{document}